\documentclass[conf]{new-aiaa}
\usepackage[utf8]{inputenc}

\usepackage{graphicx}
\usepackage{amsmath}
\usepackage[version=4]{mhchem}
\usepackage{siunitx}
\usepackage{longtable,tabularx}
\usepackage{acro}
\usepackage{svg}

\usepackage{etoolbox}
\newtoggle{urs}
\toggletrue{urs}

\usepackage{mathtools}
 \usepackage{todonotes}
\usepackage{tikz}
\usepackage{pgfplots}
\usepackage{subcaption}
\usetikzlibrary{intersections,shapes.arrows}
\usetikzlibrary {arrows.meta} 
\usepackage{float}
\usepackage{matlab-prettifier}
\usepackage{algpseudocode}
\pgfplotsset{compat=newest}
\usetikzlibrary{calc}
\usepackage{multirow,multicol}
\usepackage{amsmath,amsfonts}
\usepackage{pgfplots}
\usepackage{siunitx}
\usepackage{ulem}
\usepackage{subcaption}
\usepackage{comment}
\pgfplotsset{width=7cm,compat=1.8}
\pgfplotsset{%
  colormap={whitered}{
  color(0cm)=(white); 
  color(1cm)=(cyan)
  }
}

\DeclareAcronym{DARE}{
  short=DARE,
  long=Deep-space Autonomous Robotic Explorer,
}

\DeclareAcronym{MuSCAT}{
  short=MuSCAT,
  long=Multi-Spacecraft Concept and Autonomy Tool,
}

\DeclareAcronym{DAREabst}{
  short=DARE,
  long=Deep-space Autonomous Robotic Explorer,
}

\DeclareAcronym{MuSCATabst}{
  short=MuSCAT,
  long=Multi-Spacecraft Concept and Autonomy Tool,
}

\DeclareAcronym{CADRE}{
 short = CADRE, 
 long = Cooperative Autonomous Distributed Robotic Exploration, 
}

\DeclareAcronym{EKF}{
  short=EKF,
  long=Extended Kalman Filter,
}

\DeclareAcronym{NEOs}{
short = NEOs, 
long = Near-Earth Objects
}

\DeclareAcronym{OREX}{
short = OSIRIS-REx,
long = {Origins, Spectral Interpretation, Resource Identification, and Security-Regolith Explorer}
}

\DeclareAcronym{SoC}{
  short=SoC,
  long= State-of-Charge, 
}

\DeclareAcronym{SRP}{
  short=SRP,
  long=Solar Radiation Pressure,
}

\DeclareAcronym{TCMs}{
    short = TCMs, 
    long = Trajectory Correction Maneuvers,
}

\DeclareAcronym{BBF}{
    short = BBF, 
    long = Bennu Body-fixed Frame,
}

\DeclareAcronym{ADCS}{
    short = ADCS, 
    long = Attitude Determination and Control System,
}

\DeclareAcronym{QQ}{   
    short = QQ,
    long = Quantile-Quantile,
}

\DeclareMathOperator*{\argmax}{argmax} 

\newcommand{\eg}{{e.g.}}
\newcommand{\ie}{{i.e.}}

\title{Autonomy in the Real-World: Autonomous Trajectory Planning for Asteroid Reconnaissance via Stochastic Optimization}

\author[a,b]{Kazuya Echigo~\footnote{Intern at Jet Propulsion Laboratory, California Institute of Technology. Doctoral Student, William E. Boeing Department of Aeronautics and Astronautics, University of Washington. \textit{Corresponding author email: }{\tt\footnotesize kazuyae@uw.edu}.}}
\author[a]{Abhishek Cauligi~\footnote{Robotics Technologist, Jet Propulsion Laboratory, California Institute of Technology.}}
\author[a]{Saptarshi Bandyopadhyay~\footnote{Robotics Technologist, Jet Propulsion Laboratory, California Institute of Technology.}}
\author[a]{Dan Scharf~\footnote{Guidance and Control Engineer, Jet Propulsion Laboratory, California Institute of Technology.}}
\author[a]{Gregory Lantoine~\footnote{Mission Design Engineer, Jet Propulsion Laboratory, California Institute of Technology.}} 
\author[b]{\\ Beh\c{c}et A\c{c}{\i}kme\c{s}e~\footnote{AIAA Fellow. Professor, William E. Boeing Department of Aeronautics and Astronautics, University of Washington.}}
\author[a]{Issa Nesnas~\footnote{Principal Robotics Technologist, Jet Propulsion Laboratory, California Institute of Technology.}}
\affil[a]{NASA Jet Propulsion Laboratory, California Institute of Technology, 4800 Oak Grove Dr, Pasadena, CA 91109 USA}
\affil[b]{William E. Boeing Department of Aeronautics and Astronautics, University of Washington, Seattle, WA 98195, USA.\\}

\begin{document}

\maketitle

\begin{abstract}
This paper presents the development and evaluation of an optimization-based autonomous trajectory planning algorithm for the asteroid reconnaissance phase of a deep-space exploration mission.~\textcolor{black}{The reconnaissance phase is a low-altitude flyby to collect detailed information around a potential landing site.}
Although such autonomous deep-space exploration missions have garnered considerable interest recently, state-of-the-practice in trajectory design involves a time-intensive ground-based open-loop process that forward propagates multiple trajectories with a range of initial conditions and parameters to account for uncertainties in spacecraft knowledge and actuation.  
%
In this work, we introduce a stochastic trajectory optimization-based approach to generate trajectories that satisfy both the mission and spacecraft safety constraints during the reconnaissance phase of the~\ac{DAREabst} mission concept, which seeks to travel to and explore a near-Earth object autonomously, with minimal ground intervention.
We first use the~\ac{MuSCATabst} simulation framework to rigorously validate the underlying modeling assumptions for our trajectory planner and then propose a method to transform this stochastic optimal control problem into a deterministic one tailored for use with an off-the-shelf nonlinear solver.
Finally, we demonstrate the efficacy of our proposed algorithmic approach through extensive numerical experiments and show that it outperforms the state-of-the-practice benchmark used for representative missions.
\end{abstract}

%
%
\section{Introduction}~\label{sec:introduction}
\lettrine{A}{u}tonomy in space missions dates back many decades, but its use has been quite limited in scope, largely because of limited sensing, onboard computing resources, and algorithms \cite{nesnas2021autonomy}. Such use has often been short in duration and involves extensive ground operator oversight. In recent years, there has been a growing need to expand and extend the use of autonomy, as evidenced by the migration of autonomous surface navigation from an enhancing capability for the Spirit, Opportunity, and Curiosity rovers to an enabling one for the Perseverance Mars rover~\cite{VermaMaimoneEtAl2023_2}.
The importance of autonomy in carrying out far-flung space exploration has been borne out through a host of missions~\cite{StarekAcikmeseEtAl2016}.
The 2023–2032 Planetary Science and Astrobiology Decadal Survey (PSDS)~\cite{NASEM2022} underscored the need for autonomy advances for several of its recommended mission concepts. 
One class of such missions is small body exploration in which spacecraft journey for several months or years through a cruise phase before approaching and potentially landing on the surface of a small body.
Although such missions have been successfully carried out in the past, missions such as Hayabusa2 and~\ac{OREX} required almost two years to complete their proximity operations~\cite{Watanabe2017, OREx018}, greatly increasing operational costs and delaying scientific analysis.
Indeed, state-of-the-practice approaches that rely heavily on ground-in-the-loop operations will fall short for upcoming missions that call for exploring more distant, dynamic, and uncertain worlds.
For example, as interest in the exploration of small bodies increases toward the objective of understanding their populations \cite{swindle2019design,tan2018}, traditional mission planning and operations techniques would poorly scale for meeting the anticipated needs of future mission concepts. Targeting ~\ac{NEOs} as a subset of small bodies offers a compelling proving ground for developing and maturing autonomy technologies, given their proximity to Earth and hence their relative accessibility, yet they still represent truly unknown worlds whose characteristics remain poorly constrained. Recent research has been advancing capabilities toward the autonomous exploration of small bodies ~\cite{KUBOTA2003,JC32006, HockmanFrickEtAl2016, NesnasHockmanEtAl2021,villa2022autonomous, SpiridonovBuehlerEtAl2024}.

In this work, we focus our attention on the guidance and control for the reconnaissance phase of the mission and formulate a planning algorithm capable of efficiently computing a trajectory for the spacecraft.
This trajectory planner must be capable of satisfying complex missions constraints, including science and spacecraft system requirements, while safely operating in the presence of system uncertainties.
In order to tackle this problem, we solve the trajectory planning problem using stochastic trajectory optimization, a paradigm that has emerged as an popular framework for generating trajectories that must satisfy a rich set of mission constraints.
\textcolor{black}{Such trajectory optimization techniques have proven effective for on-board use in multiple flight missions, including the upcoming~\ac{CADRE} Lunar rover mission~\cite{DeLaCroixRossiEtAl2024}, and we leverage the tremendous advances in nonlinear optimization solvers in the past decade to efficiently solve a deterministic reformulation of this stochastic optimal control problem.}
Our proposed approach has potential to greatly accelerate the trajectory planning portion of most mission operation cycles, reducing the time required from the order of a few days to the order of a few minutes. 
In contrast, the state-of-practice entails significant ground-in-the-loop intervention and the manual design of trajectories that satisfy various subsystem requirements~\cite{OREx018, LevineWibbenEtAl2022}.

\begin{figure}[t]
    \centering
    \includegraphics[width = 0.8\textwidth]{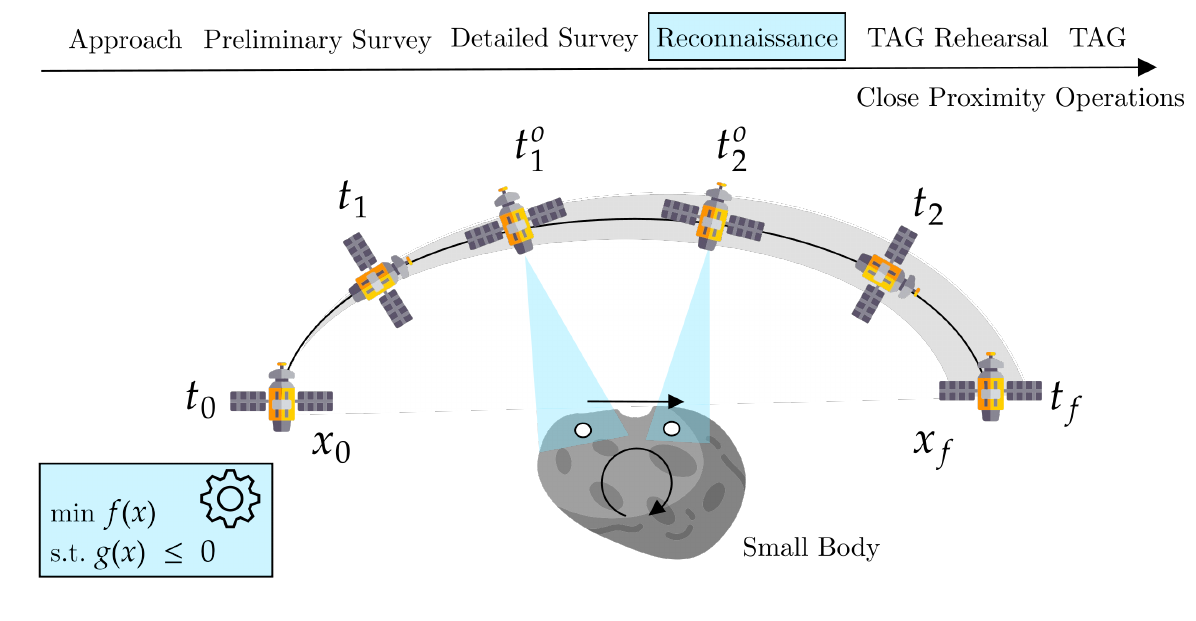}
    \caption{Overview: The proposed framework provides a trajectory for the reconnaissance phase. The realistic mission concept considered in this paper includes pre-scheduled $4$ times maneuvers ($t_0, t_1, t_2$, and $t_f$) and observations throughout the pre-scheduled observation time window (from $t_1^o$ to $t_2^o$). The proposed planner uses a stochastic optimization-based approach to efficiently deal with uncertainties and all realistic mission constraints. As a result, mission planners enjoy the efficiency, flexibility, and robustness provided by the proposed trajectory planner.}
    \label{fig:missionconcept}
\end{figure}

\textit{Statement of Contributions: }
In this paper, we present an optimization-based trajectory planning algorithm for use in the reconnaissance phase of a small body proximity operations mission.
In particular, we formulate our planner using the mission configuration for the proposed~\ac{DARE} mission~\cite{NesnasHockmanEtAl2021}, but we emphasize that our proposed approach generalizes to numerous proposed small body exploration missions.
Our contributions are listed as follows: 
\begin{itemize}
\item We validate our modeling assumptions for the trajectory planning problem by quantifying the various sources of environmental uncertainties using the~\ac{MuSCAT} simulation framework~\cite{BandyopadhyayNakkaEtAl2024}.
\item We present a stochastic optimal control problem that reconciles model fidelity with computational complexity. We then reformulate this stochastic optimal control problem into a deterministic one such that it is computationally tractable for solving with an off-the-shelf nonlinear optimization solver.
\item We demonstrate the efficacy of our propose trajectory planning approach through extensive numerical experiments and demonstrate that it outperforms state-of-the-practice approaches used in missions (\eg{},  ~\ac{OREX}).
\end{itemize}
Through this effort, we aim to demonstrate that our proposed approach would serve as a key enabling capability toward autonomous proximity operations for future small-body exploration missions.

\textit{Paper Organization: }
The remainder of this paper is organized as follows: Section~\ref{sec:related_works} reviews relevant literature in the area of spacecraft trajectory planning for small body proximity operations and recent advances in trajectory optimization-based approaches.
Section~\ref{sec:mission_modeling} reviews the~\ac{DARE} mission concept, the reconnaissance phase studied in this work, and the~\ac{MuSCAT} simulation framework.
Next, Section~\ref{sec:mission_verification} provides a numerical analysis to assess the modeling fidelity required for the trajectory planner.
We introduce the corresponding stochastic optimal control problem in Section~\ref{sec:proposed_planner}.
Section~\ref{sec:solution_algo} constitutes the main technical contribution and reformulates the stochastic optimal control problem as a deterministic one.
Finally, Section~\ref{sec:numerical_examples} presents Monte Carlo results to show the efficacy of the proposed approach and offers comparisons with the current practice used in~\ac{OREX}.


\begin{figure}[t!]
    \centering
    \includegraphics[width = 0.8\textwidth]{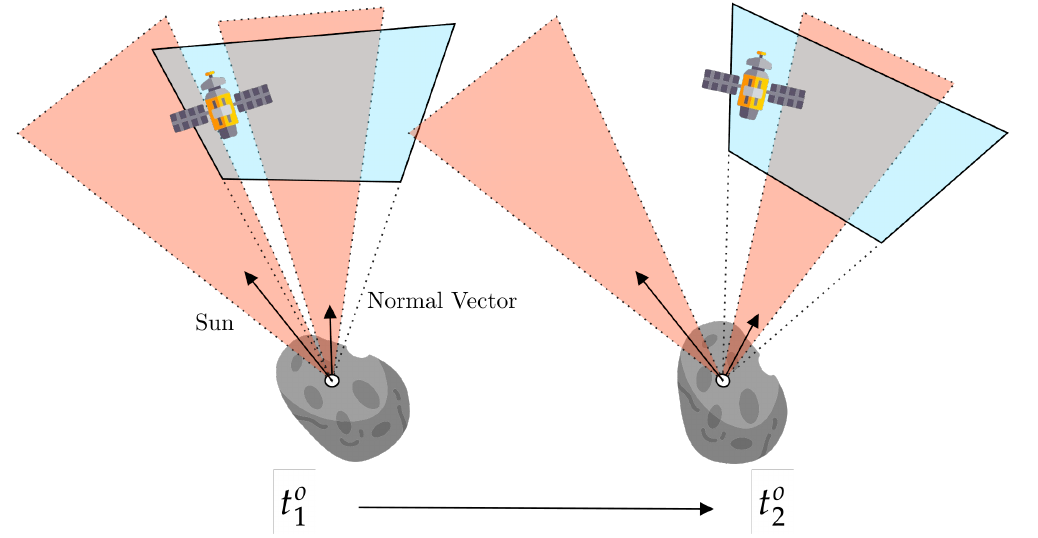}
    \caption{
   Overview of observation: The spacecraft must meet all observation constraints \textit{continuously} throughout the designated observation time window. Red regions indicate keep-out zones, whereas blue regions represent keep-in zones. Intersections of the blue and red zones are also considered keep-out zones. Those zones may move differently depending on different factors, such as the relative position of the spacecraft, the normal vector of a sample site, and the vector from the site to the Sun.
    }
    \label{fig:observationconcepts}
\end{figure}

\section{Related Work}~\label{sec:related_works}
Historically, planning for proximity operations around small bodies has involved significant manual supervision~\cite{OgawaTeruiEtAl2020, NormalMillerEtAl2022}.
More recently, autonomous trajectory planning for small body proximity operations under uncertainty has been a nascent area of research.
A commonly used approach for solving this problem has been to use sampling-based motion planning methods~\cite{StarekSchmerlingEtAl2016,FaghihiTavanaEtAl2022,DekaMcmahon2023}, wherein search-based methods are employed to solve the guidance problem.
Although such sampling-based methods can effectively find collision-free trajectories in the presence of obstacles, the computation times required scale poorly as more challenging nonlinear dynamical constraints are necessary.
Alternative approaches that have extensively been investigated include indirect optimal control techniques, which derive the necessary conditions of optimality using principles from calculus of variations~\cite{Kirk2012}.
However, indirect optimal control techniques struggle to accommodate challenging state constraints and modeling uncertainty.
\textcolor{black}{
In this work, we rely on direct methods based on nonlinear trajectory optimization techniques, which are suitable for handling these challenges. These techniques for autonomous trajectory planning have become the primary solution technique for efficiently solving such challenging optimal control problems~\cite{LiuLu2014,LiuLuEtAl2017,Kelly2017}.}
Traditionally, aerospace applications have focused on reformulating a non-convex trajectory optimization problem as a convex optimization solver and then use an off-the-shelf convex solver to efficiently solve the problem~\cite{AcikmeseCarsonEtAl2013}.
However, in the past few years, there have been tremendous advances in solving more challenging classes of non-convex trajectory optimization problems in real time~\cite{MalyutaYuEtAl2021} and several efforts have focused on developing solvers tailored for non-convex trajectory optimization problems~\cite{HowellJacksonEtAl2019, MastalliBudhirajaEtAl2020}.
Although such non-convex optimization problems lack the strong theoretical guarantees enjoyed by convex formulations, they have become a cornerstone of modern autonomy frameworks as they capture the realistic, nonlinear constraints present in most problems and have been shown to work well in practice.
For example, the upcoming~\ac{CADRE} Lunar rover mission will make use of the first on-board non-convex trajectory optimization planner to the best of the authors knowledge~\cite{DeLaCroixRossiEtAl2024}.

In this work, we rely on these aforementioned advances in non-convex trajectory optimization solvers for the problem of small body proximity operations.
The three primary sources of non-convexity for this problem are (1) the nonlinear dynamical constraints under the presence of uncertainty, (2) \textcolor{black}{collision avoidance constraints}, and (3) some observation constraints.
Tackling the problem of safe planning under uncertainty is a rich and extensive area of study.
One approach is that of robust optimization, wherein the uncertainties are considered to lie within some predefined, bounded range~\cite{BenTal2009}.
With robust optimization, trajectory planners try and solve for trajectories that are feasible under any possible realization of uncertainty that the system may encounter, but this can often lead to overly conservative results by over-assessing those worst-cases~\cite{ErnstDenHertog2020}.
\textcolor{black}{
Alternatively, works such as~\cite{OguriMcMahon2021,RizzaTopputoEtAl2024} use stochastic optimization-based approaches, which more accurately represent risk and uncertainties.}
\textcolor{black}{
In~\cite{OguriMcMahon2021}, the authors propose a generic stochastic optimization-based framework for proximity operations around small bodies. Closest to our approach, in~\cite{RizzaTopputoEtAl2024}, the authors propose a sequential convex optimization-based framework for close proximity operations, but this approach can lead to artificially infeasible solutions as the authors omit covariance as a decision variable and instead propagate the covariance using the previous solution~\cite{LahrZanelliEtAl2023}. 
In our work, we focus on the reconnaissance phase, where we identified and ensured a realistic mission scenario with representative uncertainties and constraints and with sufficient model fidelity.
}


\begin{figure}[t!]
    \centering
    \includegraphics[width=0.8\textwidth]{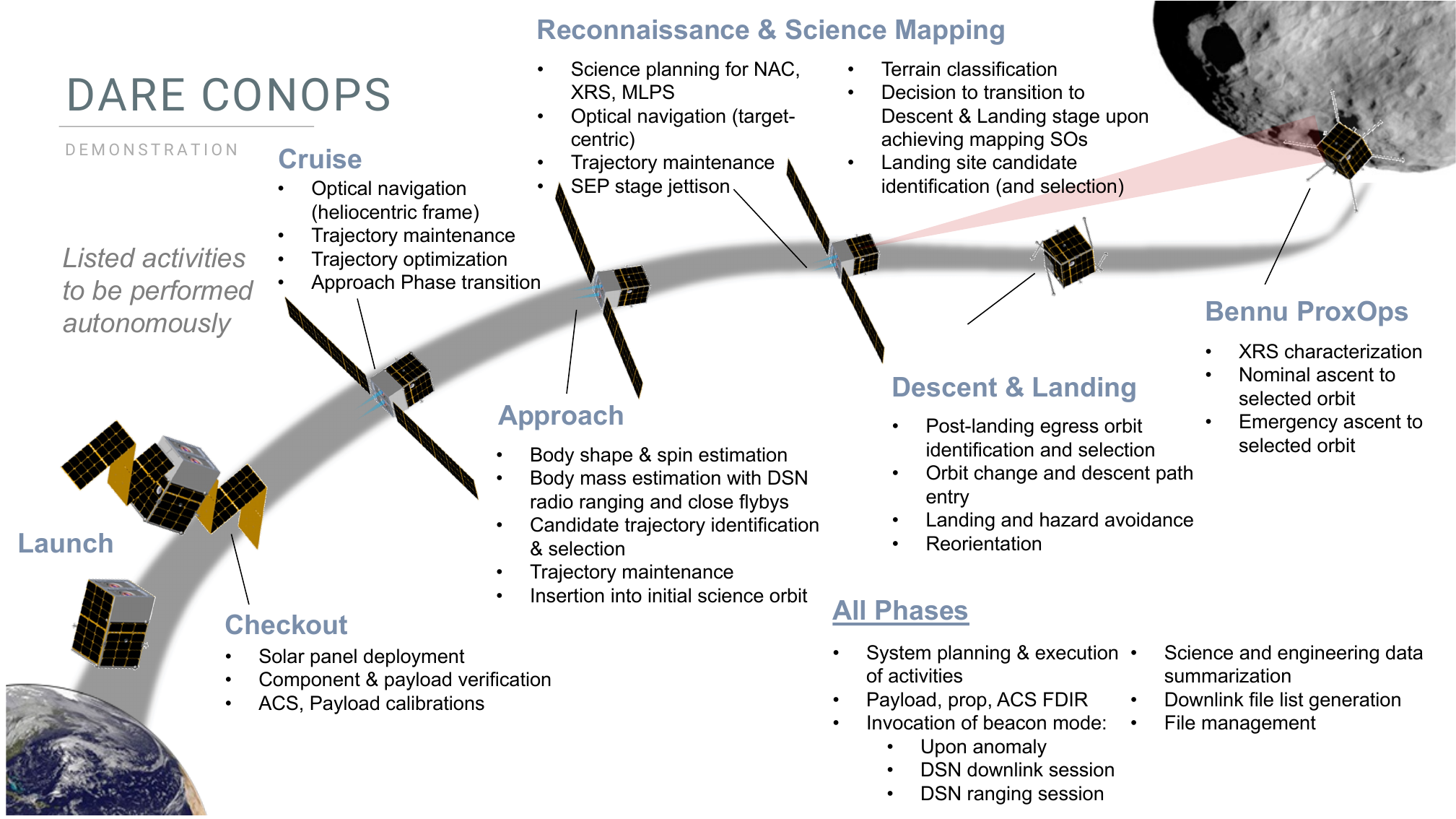}
    \caption{Overview of the Deep-space Autonomous Robotic Explorer (DARE) mission concept that seeks use autonomy across all mission phases from approach through proximity operations, landing and surface operations. In this work, we focus on the reconnaissance phase for science mapping and landing site identification.}
    \label{fig:conceptDARE}
\end{figure}
\section{Modeling Mission Concept for the Reconnaissance Phase}~\label{sec:mission_modeling}
\label{Modeling Mission Concept}
This section introduces the~\ac{DARE} mission concept and the reconnaissance phase studied in this paper. We also review the~\ac{MuSCAT} simulation framework.
To root our autonomous trajectory planning problem within the context of a deep space exploration mission, we consider the reconnaissance phase of the proposed~\ac{DARE} mission concept, whose overview is depicted in Figure~\ref{fig:conceptDARE}.
Furthermore, in order to evaluate the practical advantages of using an autonomous trajectory planner over the current practice, we design a realistic mission concept tailored to match the reconnaissance phase of the~\ac{OREX} mission~\cite{LevineWibbenEtAl2022}. For the same reason, we use Bennu as the target small body for the~\ac{DARE} mission concept.

\subsection{Overview of the Mission}
Figure~\ref{fig:missionconcept} illustrates the overall structure of the mission concept.
After the approach phase, during which the spacecraft estimates the physical parameters of the small body (e.g., rotation rate, rotation axis, and refined relative orbit), it starts the reconnaissance phase to gather more detailed information about the body including its surface topography to identify candidate landing sites~\cite{NesnasHockmanEtAl2021}.
To identify the most suitable site, the spacecraft conducts low-altitude reconnaissance over each candidate landing site and captures images from multiple altitudes.
The spacecraft starts the reconnaissance phase from a safe home orbit or position, followed by a flyby, and aims to return to the safe orbit or position at the end. For example, the reconnaissance phase for~\ac{OREX} began from a $\SI{1}{km}$ circular safe home orbit~\cite{LevineWibbenEtAl2022}. However, different asteroid shapes and spacecraft models may employ alternative approaches (\eg{}, Hayabusa2 hovered above Ryugu, whose hovering point was called the home position~\cite{Kikuchi2020}). 
The proposed planner is designed to handle these variations by adjusting the boundary conditions in its internal optimization problem, ensuring flexibility to accommodate different mission requirements.
The reconnaissance scenario here includes four~\ac{TCMs} as these are the maximum number of maneuvers the conceptual model could execute within the reconnaissance phase timespan,~\textcolor{black}{considering the limitations imposed by the time and battery consumption required for delta-V maneuver.}
Each reconnaissance orbit observes a single landing site over a defined period for two reasons:
\begin{enumerate}
    \item The reconnaissance phase is crucial for selecting landing sites, so missions typically allocate sufficient time to this phase. As a result, this approach allows the spacecraft to follow a simple trajectory without the unnecessary complexity, such as observing multiple landing sites at one time. For instance, in the case of~\ac{OREX}~\cite{LevineWibbenEtAl2022}, the entire process, from planning to executing all observation trajectories, took several months.
    \item Due to the collision risks associated with a low-altitude reconnaissance orbit, our approach favors simple orbital shapes that allow for safer execution.
\end{enumerate}
During the reconnaissance phase, except for maneuvers and observation, the spacecraft attitude is controlled to face the solar arrays toward the Sun.
Moreover, the reconnaissance orbit design must meet several observation constraints (\eg{} a constraint on the angle of the sun and the angle of the spacecraft camera relative to the surface normal of the landing site) throughout the prescribed observation time window, as depicted in Figure~\ref{fig:observationconcepts}.

\subsection{The Deep-space Autonomous Robotic Explorer (DARE) Mission Concept}

\begin{figure}[t!]
    \centering
    \includegraphics[width=0.6\textwidth]{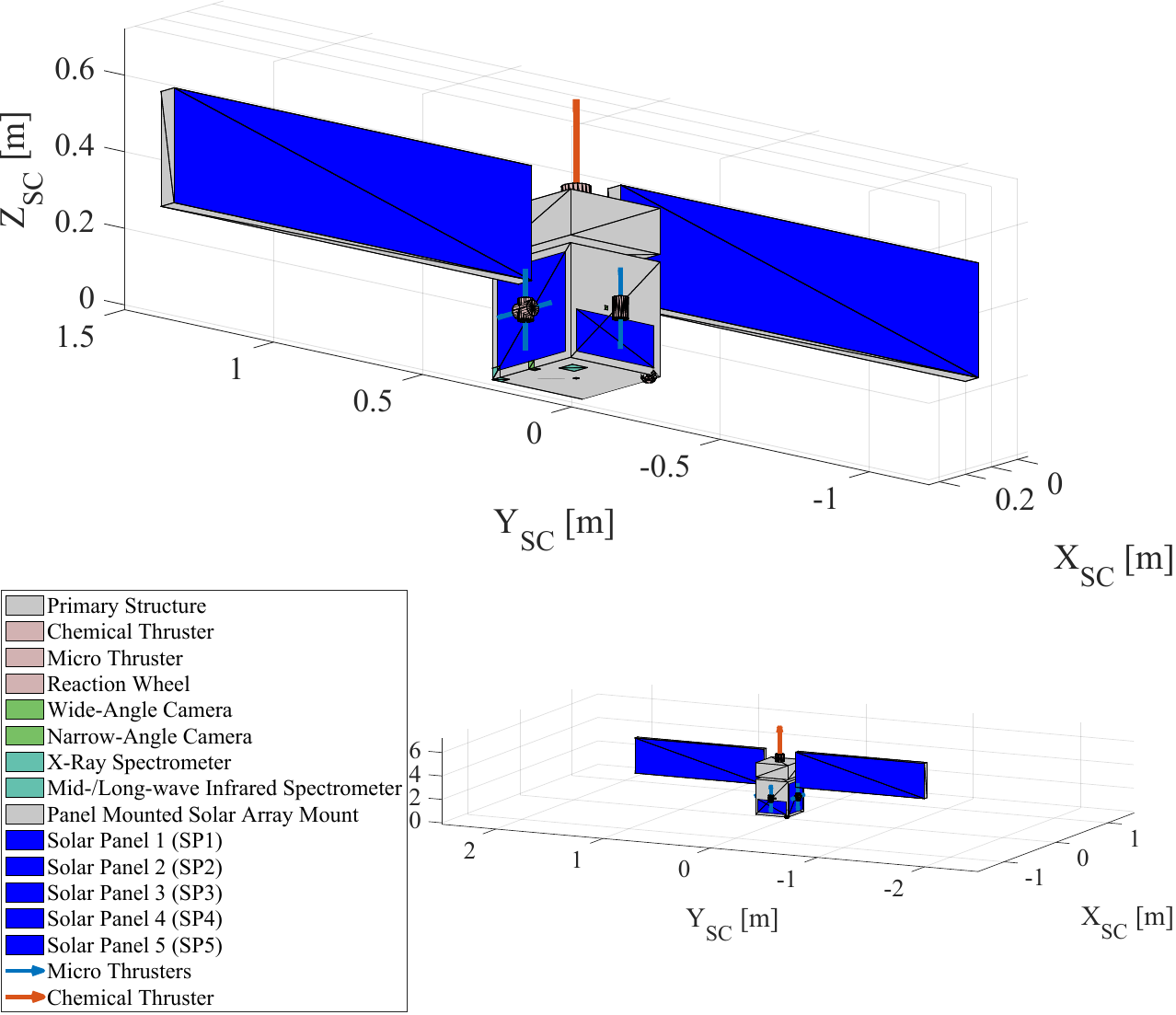}
    \caption{The design of a conceptual spacecraft for the~\ac{DARE} project.}
    \label{fig:shapeDARE}
\end{figure}

In order to design a practical trajectory planner, we use a conceptual SmallSat spacecraft design from the~\ac{DARE} project as depicted in Figure~\ref{fig:conceptDARE}~\cite{NesnasHockmanEtAl2021}.  
\ac{DARE}
is a mission concept seeking to advance mission-life-cycle autonomy. DARE uses the exploration of near-Earth objects as a means to advance autonomy in a real space environment, serving as a stepping stone toward the exploration of more remote and extreme target destinations such as ocean worlds. The mission concept targets~\ac{NEOs} because they represent unexplored bodies whose shape, gravity, and rotations are unknown \textit{a priori}. Further, several~\ac{NEOs} could be rendezvoused with and landed on or touched by small satellites ($< \SI{180}{kg}$).~\ac{NEOs} are well-suited targets for autonomy because they are abundant, are representative of more extreme destinations with large uncertainties and dynamic interactions with a spacecraft. 
Their low gravity and hence slow dynamics provide some relief that allow more time for the onboard autonomy to make decisions and take actions. Moreover, the low forces of spacecraft contact with the body reduce risk of damage.
This is necessary for an endeavor that is aiming to make significant advances to spacecraft autonomy. 

As seen in Figure~\ref{fig:shapeDARE}, this $\SI{0.3}{m} \times \SI{3.0}{m} \times \SI{0.6}{m}$ spacecraft consists of two stages: (a) a propulsion stage with a fixed chemical thruster\footnote{This is a departure from DARE, which baselined solar-electric propulsion engine for its larger delta V} and two large one-time unfoldable solar arrays and (b) the main stage with eight cold-gas micro-thrusters and reaction wheels for attitude control (see Table~\ref{tab:uncertaintyquantification} for the nominal components used in our modeling). The main stage also includes~\ac{ADCS} component (IMU, star tracker, sun sensor) for state estimation and telecom (with a phase-array antenna mounted on one side of the body).
The spacecraft generates power from five solar panels with two large arrays mounted on the propulsion stage and three smaller ones on the sides of the main stage.
Throughout operations, the power generated by these panels is used directly for the spacecraft's active components and the excess power is stored in two \SI{40}{Wh} lithium-ion batteries.

\begin{figure}
    \centering
    \includegraphics[width = 0.8\textwidth]{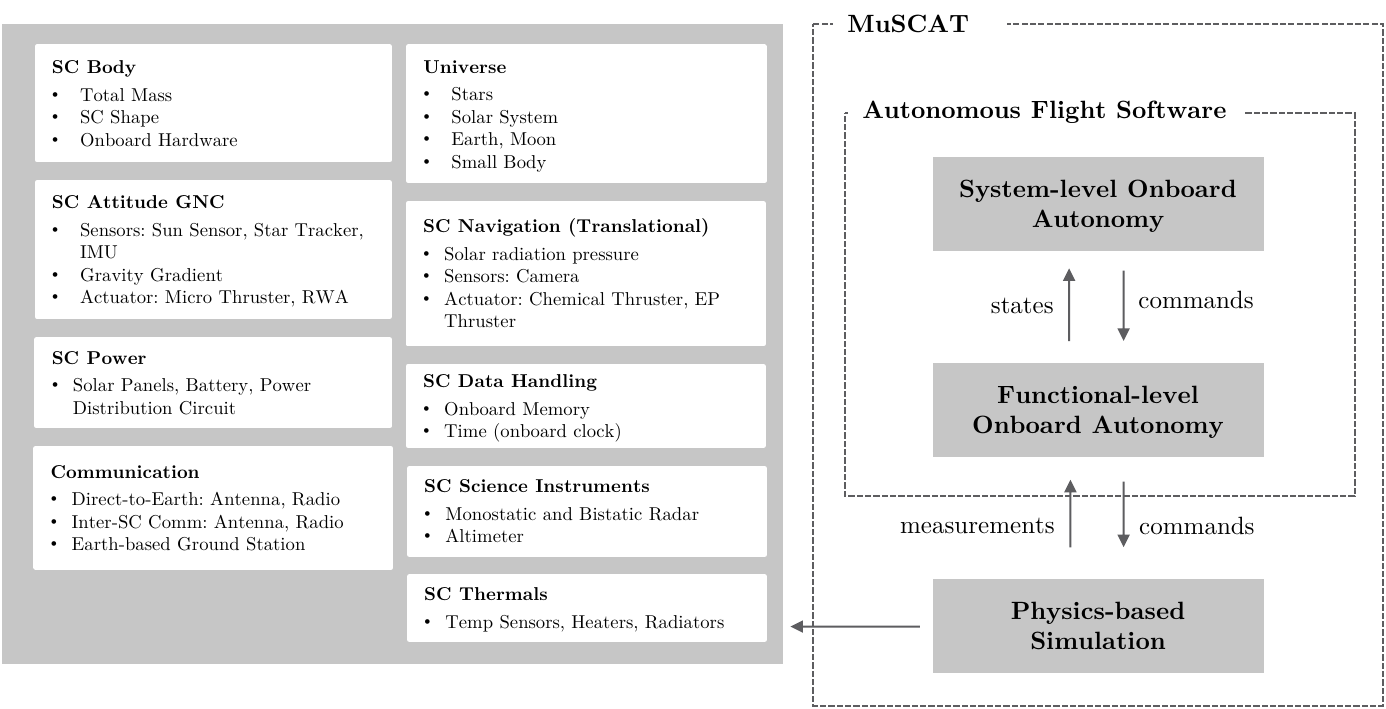}
    \caption{Overview of Multi-Spacecraft Concept and Autonomy Tool (MuSCAT). Figure shows the different components of the spacecraft that~\ac{MuSCAT} considers. ~\ac{MuSCAT} serves as a test bed for understanding which mission process may benefit from autonomy.}
    \label{fig:conceptMuSCAT}
\end{figure}

\subsection{The Multi-Spacecraft Concept and Autonomy Tool (MuSCAT)}~\label{subsect:muscat}
We evaluate the efficacy of the proposed method using~\ac{MuSCAT}, which is a software for an end-to-end mission concept simulation to test autonomy~\cite{BandyopadhyayNakkaEtAl2024,AgiaVilaEtAl2024}.
The~\ac{MuSCAT} framework is an open source software tool developed at JPL~\footnote{The code of the simulation is available at~\url{https://github.com/nasa/muscat}.} and provides an integrated platform for testing various subsystems of spacecraft simultaneously.
As seen in Figure~\ref{fig:conceptMuSCAT},~\ac{MuSCAT} can simulate subsystems such as navigation, attitude GNC, power, communication, thermal, and science instruments.
Therefore, this tool is suitable for effectively evaluating the performance of our autonomous trajectory planning algorithm within our mission concept.
With~\ac{MuSCAT}, we will investigate the degree of model fidelity required for our proposed planner through a Bennu reconnaissance case study.

\section{Model Verification}~\label{sec:mission_verification}
In this section, we present the numerical simulations conducted in~\ac{MuSCAT} to validate the modeling assumptions that will be incorporated in our reconnaissance phase trajectory planner.
Through these simulations, we develop and formulate the appropriate modeling choices:
\begin{enumerate}
    \item The state used in our trajectory planning and, in particular, the omission of the attitude as a part of the nonlinear optimization.
    \item The deterministic (\ie{}, nominal) dynamics used to model the spacecraft translational dynamics. 
    \item The particular sources of uncertainty considered (\eg{}, initial localization, control, among others).
    \item The model used to formulate and propagate uncertainty.
\end{enumerate}
\textcolor{black}{Before executing low-altitude reconnaissance phases with the risk of collision, it is essential to have sufficiently accurate environmental information about the shape and the dynamics of the small body in practice~\cite{WilliamsAntreasianEtAl2018, Kikuchi2020}. 
Therefore, although the proposed trajectory planner is capable of handling uncertainty arising from a lack of \textit{knowledge} about the environment (\eg{}, uncertainty in the landing site position and the parameters of the gravitational model), we do not address it to ensure a direct comparison between our approach and current practice.}

\subsection{Simulation Overview}
\label{simconfig}
This section identifies the appropriate model fidelity for our proposed planner through numerical simulations imitating the low-altitude (about \SI{625}{m}) reconnaissance phase of~\ac{OREX} for the primary sample site Nightingale, which is referred to as Reconnaissance B phase. 
We implemented simulations in~\ac{MuSCAT} with the conceptual SmallSat spacecraft design from the~\ac{DARE} project.

In order to follow the Reconnaissance B phase trajectory, the simulation begins at the point where the Reconnaissance B phase begins, simultaneously executing the same delta-V burn as the~\ac{OREX} mission.
We note that we use the~\ac{OREX} SPICE Kernels to construct trajectory and environmental information for this phase~\cite{ACTON199665,ACTON20189}~\footnote{Data is available at~\url{https://naif.jpl.nasa.gov/pub/naif/pds/pds4/orex/orex_spice/document/spiceds_v001.html}.}.
\textcolor{black}{The initial attitude of the conceptual spacecraft model is pre-determined so that its chemical thruster aligns with the planned delta-V direction.}
During the ballistic flight, attitude control is performed so that the power generation is maximized.
To quantify the impact of model errors and uncertainty, we measure the dispersion at the end of the phase.
In other words, we measure the distance between the terminal position of the propagated trajectory for each sample and the terminal position of the nominal trajectory, where the nominal trajectory is generated in a system without uncertainty. Moreover, we set a dispersion threshold at $\SI{10}{m}$. If the dispersion introduced by a particular modeling approach is below this threshold, we utilize it for our proposed trajectory planner. We note that if several modeling approaches result in a dispersion of less than $\SI{10}{m}$, we select the model with the lowest fidelity among them.

\subsection{Decoupling Attitude and Battery~\ac{SoC}}~\label{subsec:decouple}
Next, we discuss the validity of decoupling attitude and the battery~\ac{SoC} from the trajectory optimization process in the proposed planner.
We take an approach where we plan a nominal trajectory only considering translational motion, with attitudes assigned afterward to the resulting trajectory. Numerical simulations reveal that this traditional and practical approach works adequately except for two cases where the spacecraft runs out of battery.
In order to address them, we additionally impose constraints in the optimization process such that the battery~\ac{SoC} is guaranteed to be above the minimum threshold.

\textit{Attitude Decoupling: }
The~\ac{ADCS} consists of multiple measurement sensors (\eg{}, star trackers and IMU sensors) and actuators (\eg{}, multiple thrusters and reaction wheels).
Although actuators used for attitude control can be considered as providing precise pointing capability, a common modeling assumption is that actuators used for translational control can inject disturbances to the system.
Under this assumption, it has been a well-established and commonly used method that assigns desired attitudes to nominal trajectories that only consider translational motion\iftoggle{urs}{}{\textcolor{black}{CITE}}. 
The widespread usage of this approach in most actual missions demonstrates that a broad class of missions is suitable for this approach\iftoggle{urs}{}{\textcolor{black}{CITE}}.
For example, the~\ac{OREX} mission took the same approach for the reconnaissance phase and modeled attitude sequence as values relative to a nadir, which was estimated by the latest onboard ephemeris data~\cite{WilliamsAntreasianEtAl2018}.
Therefore, uncertainty in science instruments pointing was directly driven by uncertainty in translational dynamics, thus the performance of translational nominal trajectory was considered more important~\cite{LevineWibbenEtAl2022}.
Finally, misalignment in the translational thruster, caused by uncertainty in the attitude control, can also be considered as part of the uncertainty in delta-V. Consequently, we conclude that we can reasonably decouple attitude dynamics from the trajectory optimization process in the planner.

\begin{figure}[t!]
     \centering
     \begin{subfigure}[b]{0.6\textwidth}
         \centering
    \includegraphics[width = 0.9\textwidth]{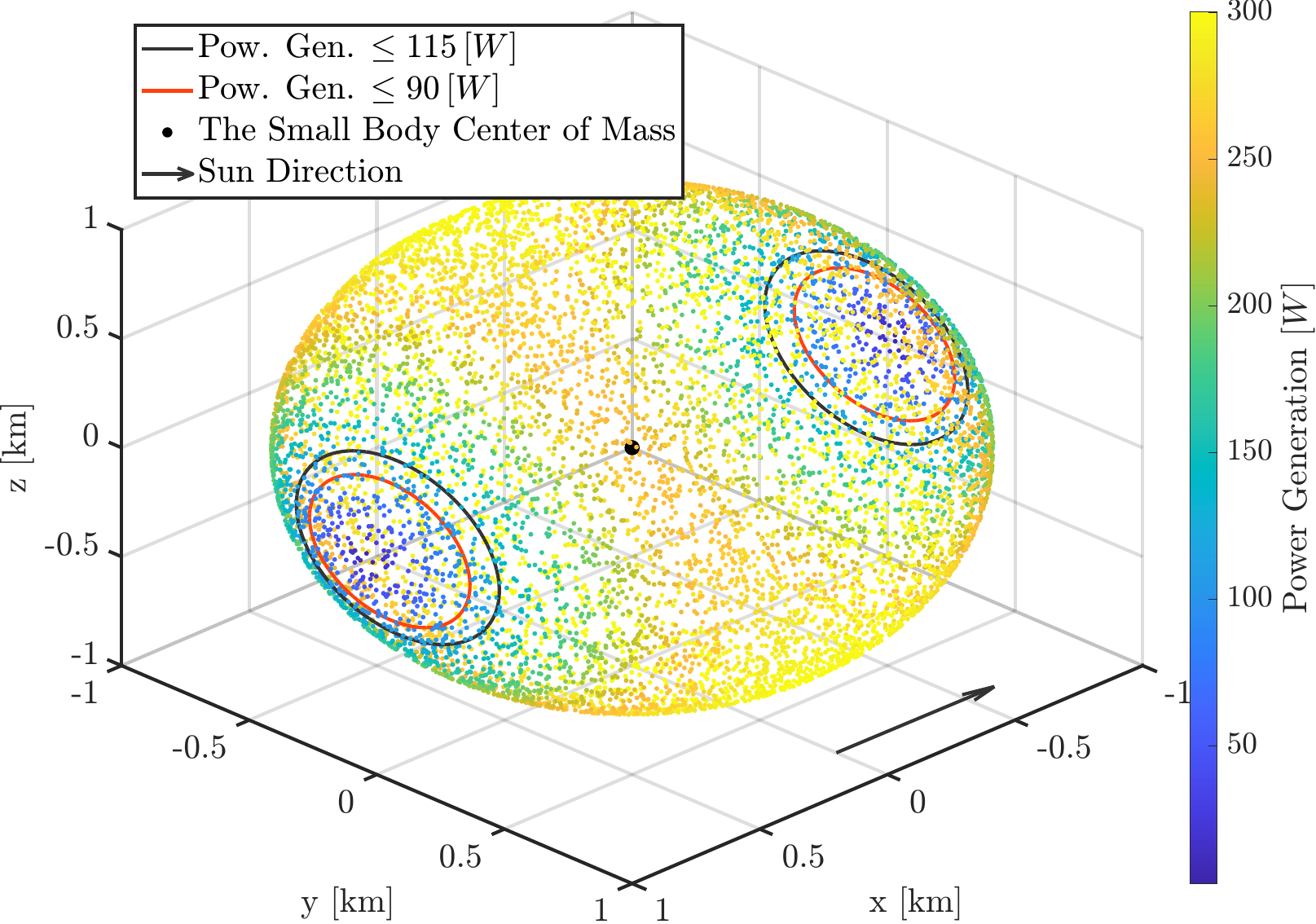}
         \caption{}
         \label{fig:firingMC1}
     \end{subfigure}
     \hfill
     \begin{subfigure}[b]{0.35\textwidth}
         \centering
         \tikzset{every picture/.style={line width=0.75pt}} 

\begin{tikzpicture}[x=0.75pt,y=0.75pt,yscale=-1,xscale=1, scale = 0.4]

\draw  [draw opacity=0] (176.64,290.44) .. controls (163.88,295.07) and (144.2,273.66) .. (132.63,242.55) .. controls (121.04,211.37) and (121.98,182.25) .. (134.73,177.51) .. controls (134.77,177.49) and (134.81,177.48) .. (134.85,177.46) -- (155.72,233.96) -- cycle ; \draw  [color={rgb, 255:red, 255; green, 13; blue, 13 }  ,draw opacity=1 ] (176.64,290.44) .. controls (163.88,295.07) and (144.2,273.66) .. (132.63,242.55) .. controls (121.04,211.37) and (121.98,182.25) .. (134.73,177.51) .. controls (134.77,177.49) and (134.81,177.48) .. (134.85,177.46) ;  
\draw  [draw opacity=0] (134.66,177.61) .. controls (149.97,172.01) and (171.73,192.65) .. (183.3,223.78) .. controls (194.9,254.96) and (191.88,284.85) .. (176.57,290.54) .. controls (176.53,290.56) and (176.49,290.57) .. (176.45,290.59) -- (155.57,234.09) -- cycle ; \draw  [color={rgb, 255:red, 255; green, 13; blue, 13 }  ,draw opacity=1 ] (134.66,177.61) .. controls (149.97,172.01) and (171.73,192.65) .. (183.3,223.78) .. controls (194.9,254.96) and (191.88,284.85) .. (176.57,290.54) .. controls (176.53,290.56) and (176.49,290.57) .. (176.45,290.59) ;  
\draw  [fill={rgb, 255:red, 255; green, 255; blue, 255 }  ,fill opacity=1 ] (500.45,104.58) .. controls (501.53,102.72) and (504.11,102.2) .. (506.23,103.42) .. controls (508.34,104.64) and (509.19,107.14) .. (508.11,109) .. controls (507.04,110.86) and (504.46,111.38) .. (502.34,110.16) .. controls (500.23,108.93) and (499.38,106.44) .. (500.45,104.58) -- cycle ;
\draw  [dash pattern={on 0.84pt off 2.51pt}]  (335.62,167.42) -- (502.16,107.66) -- (546.67,93.33) ;
\draw [shift={(333.41,168.21)}, rotate = 340.26] [color={rgb, 255:red, 0; green, 0; blue, 0 }  ][line width=0.75]      (0, 0) circle [x radius= 3.35, y radius= 3.35]   ;
\draw  [draw opacity=0] (191.68,329.95) .. controls (166.14,339.27) and (129.48,303.86) .. (109.72,250.72) .. controls (89.93,197.5) and (94.59,146.65) .. (120.13,137.15) .. controls (120.23,137.12) and (120.32,137.08) .. (120.42,137.05) -- (155.97,233.52) -- cycle ; \draw  [color={rgb, 255:red, 33; green, 33; blue, 33 }  ,draw opacity=1 ] (191.68,329.95) .. controls (166.14,339.27) and (129.48,303.86) .. (109.72,250.72) .. controls (89.93,197.5) and (94.59,146.65) .. (120.13,137.15) .. controls (120.23,137.12) and (120.32,137.08) .. (120.42,137.05) ;  
\draw  [draw opacity=0] (120.42,137.05) .. controls (145.96,127.72) and (182.62,163.14) .. (202.38,216.27) .. controls (222.17,269.49) and (217.51,320.34) .. (191.97,329.84) .. controls (191.87,329.88) and (191.78,329.91) .. (191.68,329.95) -- (156.13,233.47) -- cycle ; \draw  [color={rgb, 255:red, 33; green, 33; blue, 33 }  ,draw opacity=1 ] (120.42,137.05) .. controls (145.96,127.72) and (182.62,163.14) .. (202.38,216.27) .. controls (222.17,269.49) and (217.51,320.34) .. (191.97,329.84) .. controls (191.87,329.88) and (191.78,329.91) .. (191.68,329.95) ;  
\draw    (203.45,320.9) -- (333.41,168.21) ;
\draw [color={rgb, 255:red, 255; green, 65; blue, 13 }  ,draw opacity=1 ] [dash pattern={on 0.84pt off 2.51pt}]  (183.45,286.12) -- (333.41,168.21) ;
\draw [color={rgb, 255:red, 255; green, 65; blue, 13 }  ,draw opacity=1 ] [dash pattern={on 0.84pt off 2.51pt}]  (134.85,177.46) -- (333.41,168.21) ;
\draw  [dash pattern={on 0.84pt off 2.51pt}]  (155.57,234.09) -- (333.41,168.21) ;
\draw    (85.69,260.07) -- (123.22,246.12) -- (155.57,234.09) ;
\draw [shift={(155.57,234.09)}, rotate = 339.61] [color={rgb, 255:red, 0; green, 0; blue, 0 }  ][fill={rgb, 255:red, 0; green, 0; blue, 0 }  ][line width=0.75]      (0, 0) circle [x radius= 3.35, y radius= 3.35]   ;
\draw    (333.49,168.43) -- (536.36,201.52) ;
\draw    (333.41,168.21) -- (463.36,15.52) ;
\draw  [draw opacity=0] (463.36,15.52) .. controls (463.9,15.28) and (464.45,15.05) .. (465,14.84) .. controls (490.54,5.35) and (526.79,39.44) .. (545.97,91.01) .. controls (565.14,142.57) and (559.98,192.07) .. (534.44,201.56) .. controls (534.35,201.6) and (534.25,201.64) .. (534.15,201.67) -- (499.72,108.2) -- cycle ; \draw  [color={rgb, 255:red, 33; green, 33; blue, 33 }  ,draw opacity=1 ] (463.36,15.52) .. controls (463.9,15.28) and (464.45,15.05) .. (465,14.84) .. controls (490.54,5.35) and (526.79,39.44) .. (545.97,91.01) .. controls (565.14,142.57) and (559.98,192.07) .. (534.44,201.56) .. controls (534.35,201.6) and (534.25,201.64) .. (534.15,201.67) ;  
\draw  [draw opacity=0][dash pattern={on 0.84pt off 2.51pt}] (536.36,201.52) .. controls (518.89,206.9) and (489.67,170.48) .. (470.66,119.37) .. controls (452.56,70.69) and (450.18,26.03) .. (464.47,15.56) -- (502.16,107.66) -- cycle ; \draw  [color={rgb, 255:red, 33; green, 33; blue, 33 }  ,draw opacity=1 ][dash pattern={on 0.84pt off 2.51pt}] (536.36,201.52) .. controls (518.89,206.9) and (489.67,170.48) .. (470.66,119.37) .. controls (452.56,70.69) and (450.18,26.03) .. (464.47,15.56) ;  
\draw [color={rgb, 255:red, 255; green, 65; blue, 13 }  ,draw opacity=1 ] [dash pattern={on 0.84pt off 2.51pt}]  (333.41,168.21) -- (483.37,50.31) ;
\draw [color={rgb, 255:red, 255; green, 65; blue, 13 }  ,draw opacity=1 ] [dash pattern={on 0.84pt off 2.51pt}]  (333.49,168.43) -- (525.16,163.29) ;
\draw  [draw opacity=0][dash pattern={on 0.84pt off 2.51pt}] (483.37,50.31) .. controls (498.68,44.71) and (520.44,65.35) .. (532.02,96.48) .. controls (543.61,127.65) and (540.6,157.55) .. (525.28,163.24) .. controls (525.24,163.26) and (525.2,163.27) .. (525.16,163.29) -- (504.28,106.79) -- cycle ; \draw  [color={rgb, 255:red, 255; green, 13; blue, 13 }  ,draw opacity=1 ][dash pattern={on 0.84pt off 2.51pt}] (483.37,50.31) .. controls (498.68,44.71) and (520.44,65.35) .. (532.02,96.48) .. controls (543.61,127.65) and (540.6,157.55) .. (525.28,163.24) .. controls (525.24,163.26) and (525.2,163.27) .. (525.16,163.29) ;  
\draw  [draw opacity=0][dash pattern={on 0.84pt off 2.51pt}] (525.16,163.29) .. controls (512.4,167.91) and (492.72,146.51) .. (481.15,115.39) .. controls (469.56,84.21) and (470.5,55.09) .. (483.25,50.35) .. controls (483.29,50.34) and (483.33,50.32) .. (483.37,50.31) -- (504.24,106.8) -- cycle ; \draw  [color={rgb, 255:red, 255; green, 13; blue, 13 }  ,draw opacity=1 ][dash pattern={on 0.84pt off 2.51pt}] (525.16,163.29) .. controls (512.4,167.91) and (492.72,146.51) .. (481.15,115.39) .. controls (469.56,84.21) and (470.5,55.09) .. (483.25,50.35) .. controls (483.29,50.34) and (483.33,50.32) .. (483.37,50.31) ;  
\draw    (130.53,135.12) -- (333.41,168.21) ;
\draw [shift={(333.41,168.21)}, rotate = 9.27] [color={rgb, 255:red, 0; green, 0; blue, 0 }  ][fill={rgb, 255:red, 0; green, 0; blue, 0 }  ][line width=0.75]      (0, 0) circle [x radius= 3.35, y radius= 3.35]   ;
\draw    (546.67,93.33) -- (609.92,70.46) ;
\draw  [draw opacity=0] (283.98,186.34) .. controls (282.01,181.12) and (280.89,175.68) .. (280.6,170.25) -- (333.41,168.21) -- cycle ; \draw  [color={rgb, 255:red, 255; green, 65; blue, 13 }  ,draw opacity=1 ] (283.98,186.34) .. controls (282.01,181.12) and (280.89,175.68) .. (280.6,170.25) ;  
\draw  [draw opacity=0] (231.99,203.09) .. controls (224.29,185.59) and (220.53,167.21) .. (220.62,149.35) -- (339.01,163.84) -- cycle ; \draw   (231.99,203.09) .. controls (224.29,185.59) and (220.53,167.21) .. (220.62,149.35) ;  
\draw    (207.45,66.5) .. controls (190.79,146.86) and (199.1,129.26) .. (216.38,166.17) ;
\draw [shift={(217.45,168.5)}, rotate = 245.77] [fill={rgb, 255:red, 0; green, 0; blue, 0 }  ][line width=0.08]  [draw opacity=0] (8.93,-4.29) -- (0,0) -- (8.93,4.29) -- cycle    ;
\draw    (321.45,275.5) .. controls (260.38,256.78) and (245.88,234.19) .. (276.04,182.86) ;
\draw [shift={(277.45,180.5)}, rotate = 121.12] [fill={rgb, 255:red, 0; green, 0; blue, 0 }  ][line width=0.08]  [draw opacity=0] (8.93,-4.29) -- (0,0) -- (8.93,4.29) -- cycle    ;
\draw    (333.49,168.43) -- (344.91,106.45) ;
\draw [shift={(345.45,103.5)}, rotate = 100.44] [fill={rgb, 255:red, 0; green, 0; blue, 0 }  ][line width=0.08]  [draw opacity=0] (8.93,-4.29) -- (0,0) -- (8.93,4.29) -- cycle    ;

\draw (331, 287.4) node [anchor=north west][inner sep=0.75pt]  [font=\large]  {$\theta _{\text{obs}}$};
\draw (184, 25.4) node [anchor=north west][inner sep=0.75pt]  [font=\large]  {$\theta _{\text{delta-V}}$};

\end{tikzpicture}
         \caption{}
         \label{fig:firingMC2}
     \end{subfigure}
        \caption{A visualization of $10,000$ Monte Carlo simulation results (a). Displayed are the maximum power generation at each sample and the boundaries where the power generation is below $\SI{115}{\watt}$ or $\SI{90}{\watt}$ (black lines and red lines, respectively). If the spacecraft position is outside of both dual cones, the battery~\ac{SoC} is above the minimum threshold (b).}
        \label{fig:firingMC}
\end{figure}
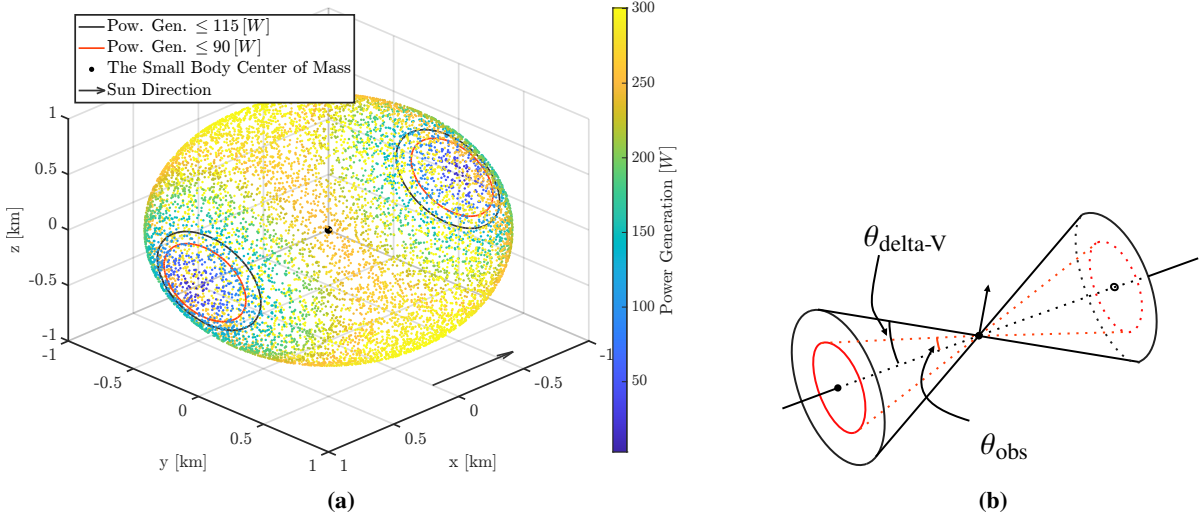

\textit{Battery~\ac{SoC} Decoupling: }
We then turn our attention to the battery~\ac{SoC} in the power subsystem. For safety reasons, the spacecraft needs to maintain the battery~\ac{SoC} above a certain level during the reconnaissance phase\iftoggle{urs}{}{\textcolor{black}{CITE. To do: ASK SAptarshi}}.
We set the lower threshold to $30\%$ and will discuss the approach of designing trajectories to ensure that the battery~\ac{SoC} does not fall below this threshold without explicitly considering it in the trajectory planner. 

During the reconnaissance phase, the attitude GNC subsystem needs to track one of the three attitude modes: 1) a delta-V mode, 2) a sun-pointing mode to maximize generated power, and 3) an observation mode.
Among those three modes, in the sun-pointing mode, the battery~\ac{SoC} never falls below the lower limit, so maintaining the battery~\ac{SoC} is not of concern.
This is because all flybys for this reconnaissance phase are designed to pass over the daytime side of the small body to meet the solar incidence angle constraints (we will also discuss later), ensuring the maximum power generation.
However, in the other two modes, commanding certain attitude slews may result in the battery~\ac{SoC} falling below the threshold.
Indeed, in the worst power generation case, there are attitudes where none of the solar panels generate power. For example, fixing the spacecraft in such an attitude to warm up the thruster during the delta-V mode could lead to running out of battery. Consequently, it is necessary to identify attitudes where the battery~\ac{SoC} is above the lower threshold and impose a state constraint to take such attitudes in the trajectory optimization process.
We will discuss the formulation of such a constraint in Section~\ref{modelstateconsts} and here we provide a numerical analysis to identify those attitudes. Since maintaining the battery~\ac{SoC} is placed at the highest priority, we leave a sufficient safety margin and identify attitudes such that power generation exceeds power consumption. 

We begin by providing details on the power consumption in the two modes, the delta-V mode and the observation mode. Using the conceptual SmallSat model, the total estimated power consumption during the thruster warm-up phase in the delta-V mode is $\SI{115}{W}$, whereas it is $\SI{90}{W}$ during the observation mode.
We note that the thruster warm-up consumes the most power in the delta-V mode.
Next, to identify attitudes where power generation exceeds these thresholds, we evaluate power generation at randomized attitudes.
Our analysis involves $10,000$ sample Monte Carlo simulations, where we place the spacecraft at random points on a sphere with a radius of $\SI{1}{km}$ centered at the Bennu center of mass. At each sampled position, we compute the attitude of the spacecraft where its bottom face towards the nadir while maximizing the generated power. We note that there remains an additional degree of freedom to consider with the angle around the axis from the small body center-of-mass to the randomized spacecraft position. With~\ac{MuSCAT}, we compute an angle with the maximum power generation.
Each sampled attitude corresponds to different actions: in the delta-V mode, it involves pointing the chemical thruster towards the direction opposite the nadir, and in the observation mode, it involves pointing the camera towards the nadir.
Figure~\ref{fig:firingMC1} shows the Monte Carlo simulation result, illustrating the maximum power generation for each sample and indicating the regions where the power generation is below $\SI{115}{W}$ or $\SI{90}{W}$. Indeed, these regions are described as dual cones (see Figure~\ref{fig:firingMC2}). The apexes of these cones are located at the center of mass, with angles $\theta_\text{obs} = \SI{18.2}{\degree}$ for the observation mode and $\theta_\text{deltaV} = \SI{23.3}{\degree}$ for the delta-V mode, respectively. In Section~\ref{sec:proposed_planner}, we will impose constraints such that the spacecraft is placed to be outside of these dual cones in order to ensure that the battery~\ac{SoC} does not fall below the lower threshold.

\begin{figure}[t!]
     \centering
     \begin{subfigure}[b]{0.45\textwidth}
         \centering
         \includegraphics[width=\textwidth]{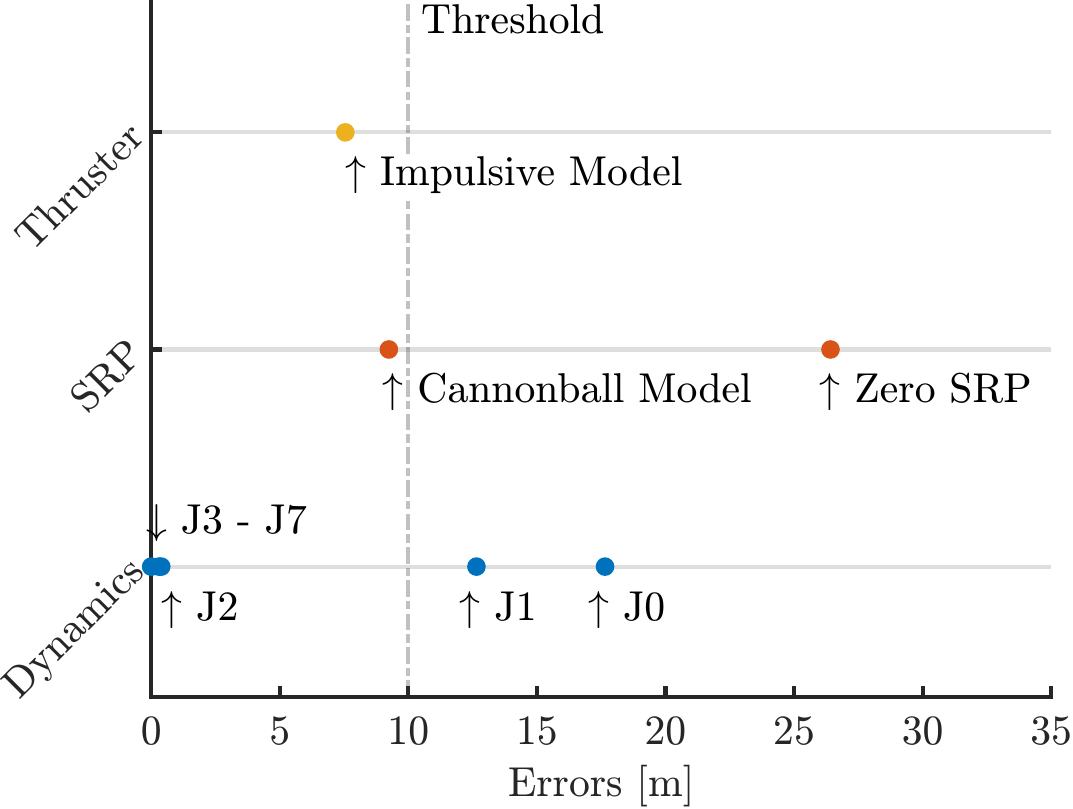}
         \caption{}
         \label{fig:model_nonuncertainty}
     \end{subfigure}
     \hfill
     \begin{subfigure}[b]{0.45\textwidth}
         \centering
         \includegraphics[width=\textwidth]{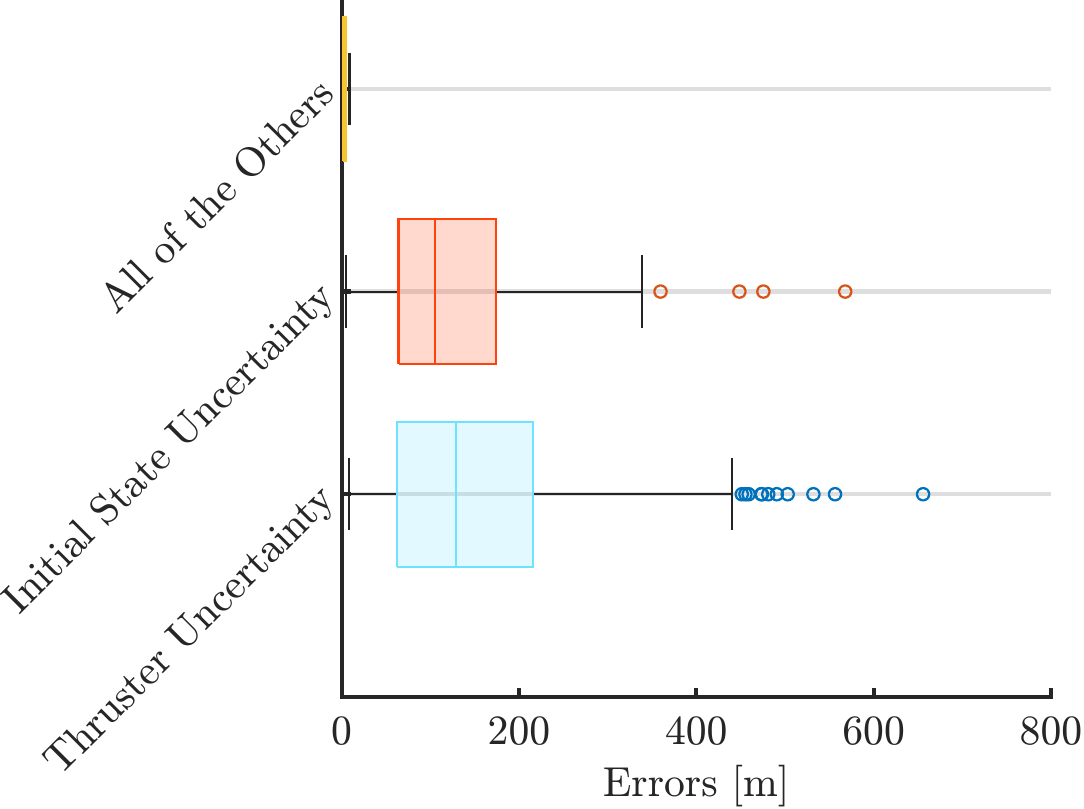}
         \caption{}
         \label{fig:model_uncertainty}
     \end{subfigure}
        \caption{Comparisons of modeling techniques and sources of uncertainty that impact the performance of the trajectory planner. Figures illustrate the dispersions from different modeling techniques (a) and sources of uncertainty (b). Figure (a) indicates that the impulsive modeling of the thruster acceleration, the cannonball modeling of the~\ac{SRP}, and the J2 modeling of perturbations are sufficiently accurate for the trajectory planner. Figure (b) highlights the significant impacts of thruster uncertainty and initial state uncertainty compared to the other sources.}
        \label{fig:two graphs}
\end{figure}

\subsection{Verifying Model Fidelity}~\label{subsec:model_assurance}
Here, we focus on verifying the appropriate model fidelity for our autonomy task. Model fidelity describes the accuracy with which a model represents the real-world system. The appropriate model fidelity balances detail against numerical tractability. We begin with the orbit propagation of the form: 
\begin{equation}
\dot{x} = f(x, \,\, t) + [0_{3\times 1}, \,\, a_\text{thrust}^\top]^\top,
\end{equation}
where a state vector $x= [r^\top, \dot{r}^\top]^\top \in \mathbb{R}^{6}$ contains a position $r = [r_x,r_y,r_z]^\top \in\mathbb{R}^3$ and a velocity $\dot{r} = [r_x,r_y,r_z]^\top \in\mathbb{R}^3$. The function $f:\mathbb{R}^{6} \times \mathbb{R} \rightarrow \mathbb{R}^{6}$ captures the autonomous system dynamics and $a_\text{thrust} \in \mathbb{R}^{3}$ is an acceleration induced by the chemical thruster of the spacecraft.
We let $x$ denote the vector relative to the small body center of mass, defined in the J2000 frame. 
In particular, we evaluated the effects of including three canonical environmental effects. These factors, typically considered in high-fidelity translational dynamics propagation, include the~\ac{SRP}, propulsion force, and gravitational force. Thus, $f$ is expressed as
\begin{equation}
   f(x, t) = [\dot{r}(t)^\top, \,\, \frac{1}{m} \{ F_\text{SRP} + F_\text{grav}\}^\top]^\top,  \label{eq:doubleintegdyn}  
\end{equation}
where $m \in \mathbb{R}$ is a mass, $F_\text{SRP}$ represents the~\ac{SRP}, and $F_\text{grav}$ is a gravitational force induced by the small body, Bennu. We will discuss the appropriate model fidelity for each force in the following sections.

\subsubsection{Solar Radiation Pressure}
In order to discuss the appropriate modeling method for the~\ac{SRP}, we evaluate the dispersion at the end of the reconnaissance phase for two approximated systems: one that does not consider~\ac{SRP}, and another that represents~\ac{SRP} using a Cannonball model.
The Cannonball model approximates the shape of the spacecraft as a sphere. We note that we consider an N-Plate model as the ground truth~\cite{JEAN2019167}. As seen in Figure~\ref{fig:model_nonuncertainty}, the dispersion due to the Cannonball model approximation was less than $\SI{10}{m}$, indicating that the Cannonball model could be reasonably accurate when the spacecraft operates for a sufficiently short duration, as in our fly-by mission.

\subsubsection{Spherical Harmonics Model}
In order to represent the gravitational force, we consider a spherical harmonics model.
This model is well known to accurately represent a gravitational model of irregularly shaped small bodies when the spacecraft is at the outside of its circumscribing sphere~\cite{Werner1996}.
As in the previous discussion, we evaluate the dispersion for eight systems by varying the order of the spherical harmonics model that represents the gravity induced by the small body, ranging from $0$ (point mass model) to $7$.
We consider an $8^\text{th}$-order model as the ground truth.
The results of the evolution of the state through the eight different systems are observed in Figure~\ref{fig:model_nonuncertainty}.
The system considering up to J2 perturbations results in terminal dispersion within $\SI{10}{m}$.
This is consistent with the observations that the spacecraft was insensitive to higher orders of the model in the~\ac{OREX} mission~\cite{Scheeres2020, Goossens2021}. Thus, we conclude that incorporating J2 perturbations in the planner adequately captures the dynamics propagation.

\subsubsection{Thruster Model}
Finally, we evaluate an impulsive model to represent the control acceleration induced by the spacecraft.
In real missions (and also in the ground truth model of~\ac{MuSCAT}), the spacecraft continues to thrust for a certain period until it achieves the desired delta-V.
On the other hand, because this thrusting period is usually much shorter than the planned time-of-flight, these control inputs are often simplified as impulse inputs using the Dirac delta functions in orbit design~\cite{Conway2010}.
To confirm the validity of this impulsive approximation, we analyze the terminal dispersion of a system where the initial thrust is approximated as an impulse.
Provided that the resulting terminal position error is $\SI{7.55}{m}$, we conclude that the impulse approximation is sufficiently accurate for this specific reconnaissance phase.

\begin{table}[t!]
        \centering
        \begin{tabular}{ccc}
        \hline
        Sources & $1 \sigma$ & Nominal Components  \\ \hline
        Sun Sensor Estimation Error& $\SI{0.05}{rad}$& SSOC-A60~\cite{SSOC-A60} \\
        Star Tracker Estimation Error & $\SI{15}{arcsec}$ & Standard Nano Star Tracker~\cite{StarTracker}\\
        IMU Sensor Estimation Error & $\SI{9.7e-5}{rad/s}$  & MASIMU01~\cite{IMU}\\
        Initial State & $3\%$ of the estimated state& -  \\
        Micro Thruster Actuation Error & $\SI{100e-6}{N}$ & $\SI{1}{mN}$ HPGP Thruster~\cite{microth} \\
        Chemical Thruster Actuation Error & $10\%$ of the nominal thrust&  $\SI{1}{N}$ HPGP Thruster~\cite{chemth}  \\
        \hline 
        \end{tabular}
     \caption{Sources of uncertainty considered in this paper.}
        \label{tab:uncertaintyquantification}
\end{table}

\begin{figure}[t!]
\begin{subfigure}{.5\textwidth}
  \centering
  \includegraphics[width = 0.9\textwidth]{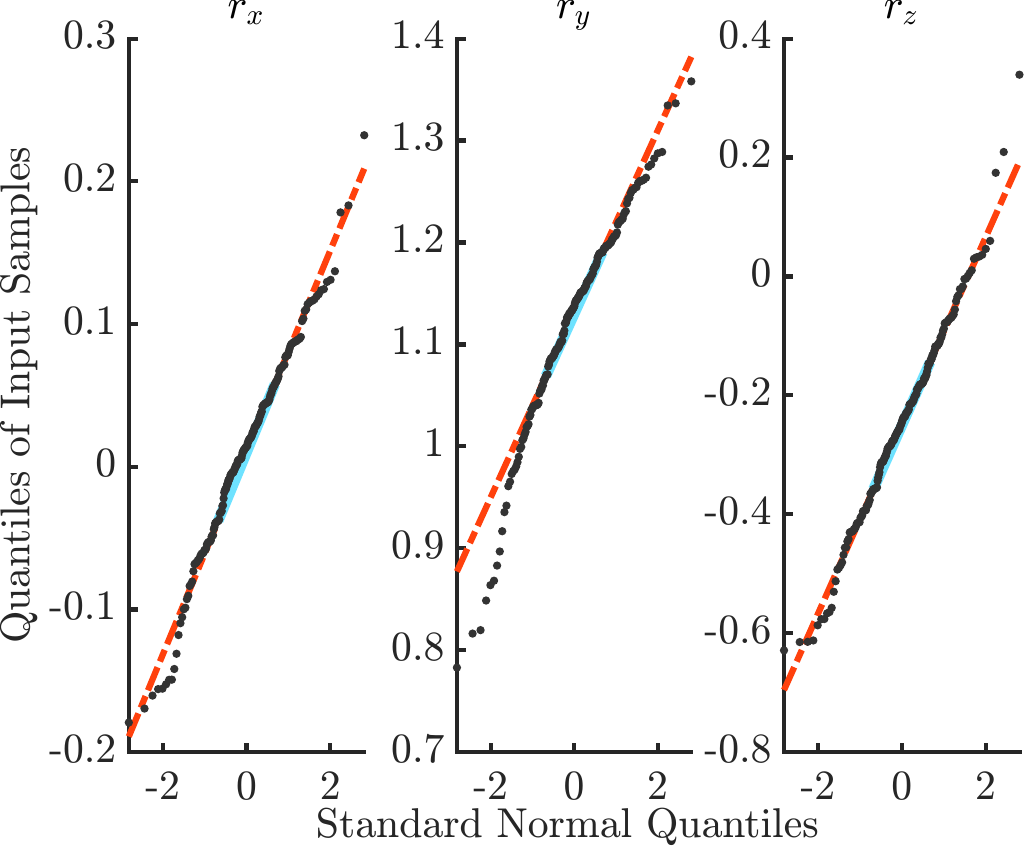}
  \caption{}
  \label{fig:qqcontrol}
\end{subfigure}
\begin{subfigure}{.5\textwidth}
  \centering
  \includegraphics[width = 0.9\textwidth]{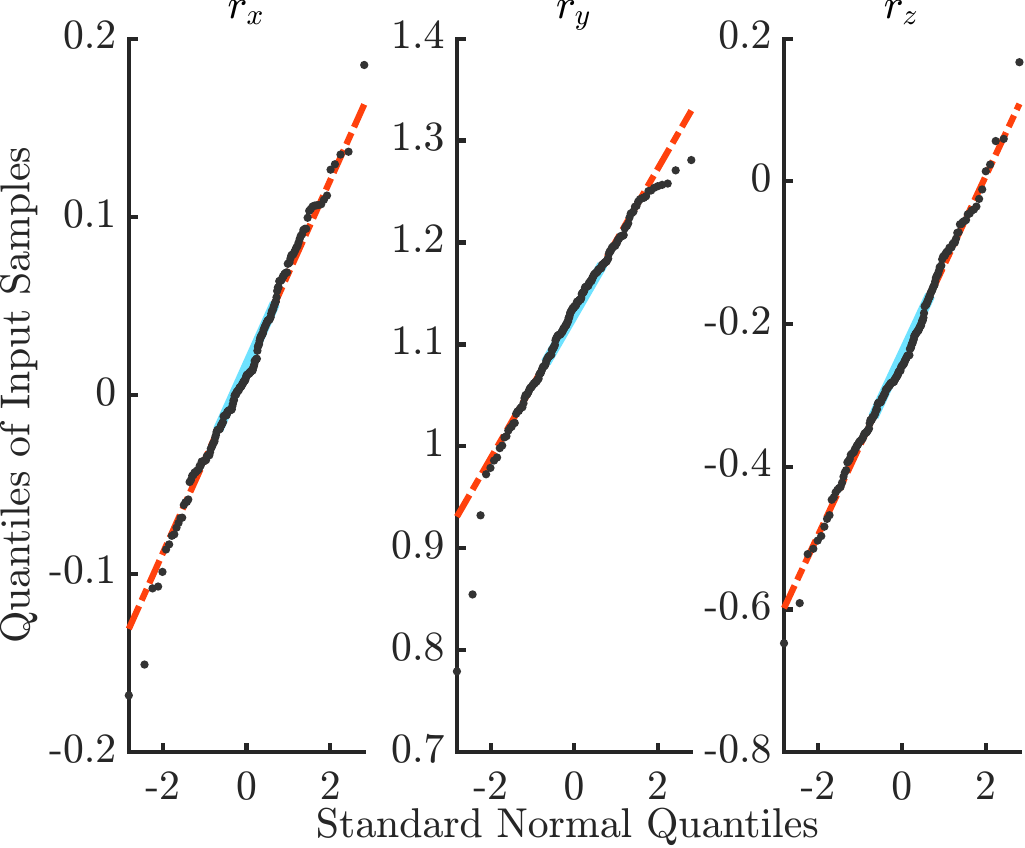}
  \caption{}
  \label{fig:qqstate}
\end{subfigure}
\caption{\ac{QQ} Plots of terminal state dispersion projected on each axis of the J2000 frame and Standard Normal. Figure (a) displays the QQ Plot of terminal state dispersion considering initial state uncertainty, whereas Figure (b) displays the QQ plot of terminal state dispersion considering thruster uncertainty. \textcolor{black}{In both figures, the black dots represent the data points of the dispersion. The red dashed line connects the first and third quantiles of those, while the blue solid line represents a segment between these two quantiles.} Given that the points in the plots approximately lie on a linear line, we conclude that the Gaussian approximation of state distributions is applicable.}
\label{fig:qqplots}
\end{figure}

\subsection{Quantifying Uncertainty and Modeling Uncertainty Propagation}~\label{subsec:model_uncertainty}
In order to evaluate sources of uncertainty, several Monte Carlo simulations are run in the~\ac{MuSCAT} simulation environment while activating each source separately. Table~\ref{tab:uncertaintyquantification} presents the sources of uncertainty considered in our simulations, specified as Gaussian distributions with zero means.
This table also details the nominal components employed in the conceptual model for the~\ac{DARE} project, all of which were simulated in the~\ac{MuSCAT} simulation.
The parameters representing uncertainty were informed by the data sheets referenced in Table~\ref{tab:uncertaintyquantification}, Monte Carlo simulations carried out for the~\ac{OREX} mission~\cite{LevineWibbenEtAl2022}, and related research on estimating uncertainty as discussed in~\cite{Bottiglieri2023}. The sources of uncertainty in the attitude dynamics include sun sensor estimation error, star tracker estimation error, IMU sensor estimation error, and micro thruster actuation error, whereas the remaining in Table~\ref{tab:uncertaintyquantification} are the sources of uncertainty in the translational dynamics. Uncertainty in the attitude GNC subsystem affects the accuracy of the maneuver and may also put additional forces in the translational dynamics during the sun-pointing mode.

As shown in~Figure~\ref{fig:model_uncertainty}, the key findings from our simulations are that a state estimation uncertainty and a thruster uncertainty are the main sources of dispersion.
We also note that attitude error from any of sources of uncertainty at the end was negligibly small (\ie{}, less than $0.1\%$ of the desired attitude at the end).
These results motivate our approach to decouple attitude dynamics from the trajectory planner. 

Next, in order to identify the appropriate technique to capture uncertainty and model its propagation within the trajectory optimization process, we assess the closeness of state distributions to normality.
Figure~\ref{fig:qqplots} displays~\ac{QQ} plots of each element of state data at the end.
A~\ac{QQ} plot is a graphical tool used to assess whether a given data follows a particular probability distribution, such as the normal distribution.
In a~\ac{QQ} plot, the quantiles of the dataset are plotted against the quantiles of a theoretical distribution.
If the dataset closely follows the theoretical distribution, then the points on the~\ac{QQ} plot will fall approximately along a straight line.
As the data in Figure~\ref{fig:qqplots} show, all points are roughly aligned along the straight line and this result indicates the validity of imposing a Gaussian assumption on the state to model the uncertainty propagation.

\begin{figure}[t!]
\begin{subfigure}{.5\textwidth}
  \centering
  \includegraphics[width = 0.9\textwidth]{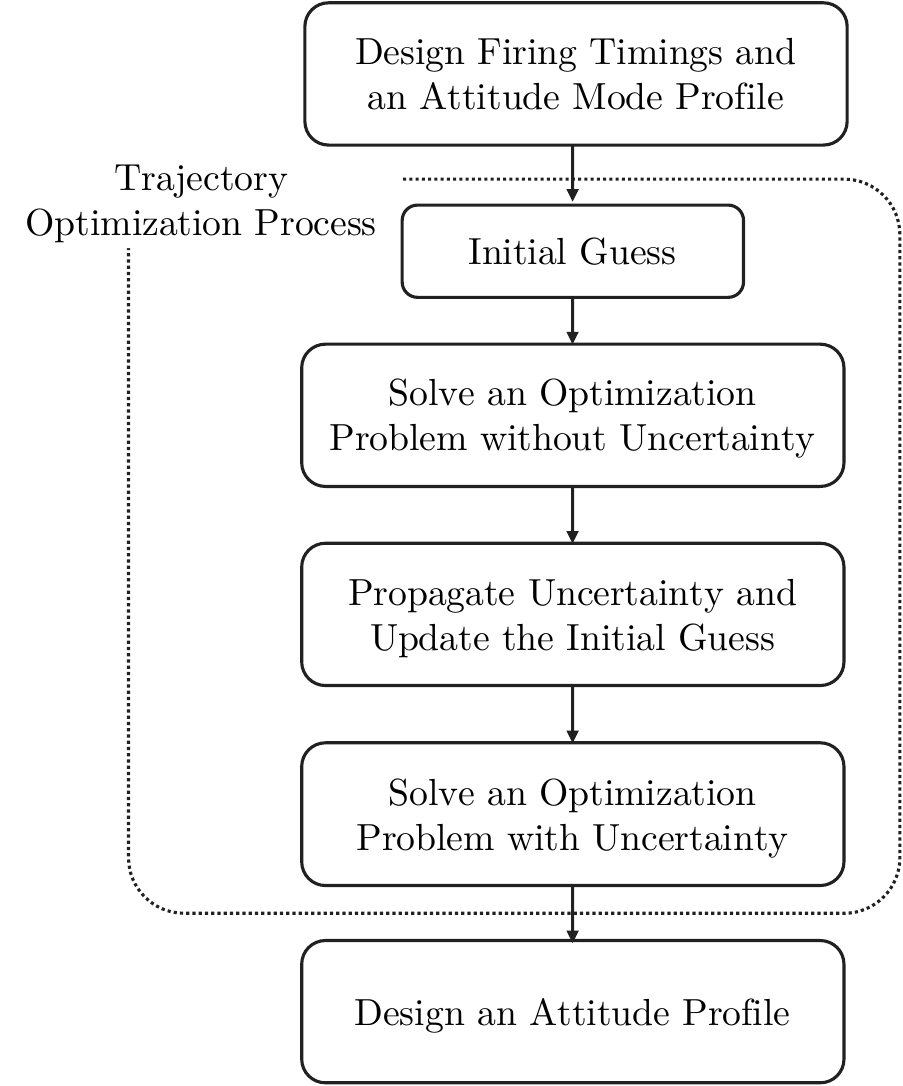}
  \caption{}
  \label{fig:schematics}
\end{subfigure}
\begin{subfigure}{.5\textwidth}
  \centering
  \includegraphics[width = 0.9\textwidth]{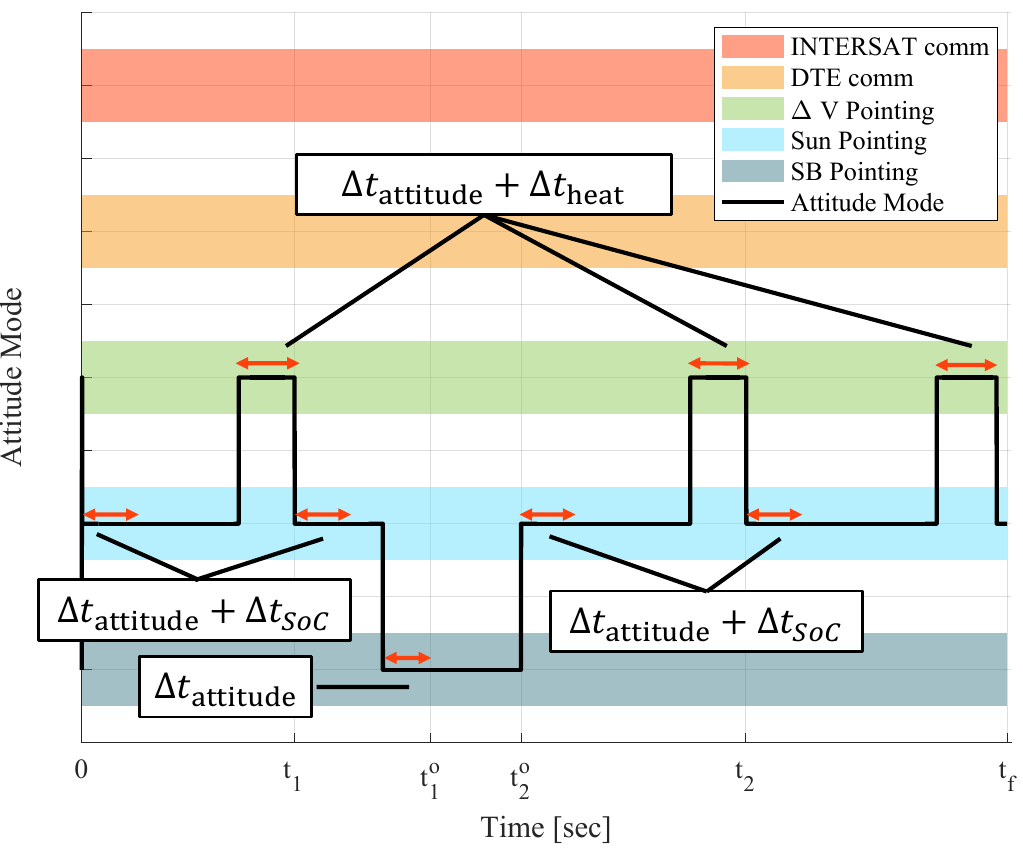}
  \caption{}
  \label{fig:attitude assign}
\end{subfigure}
\caption{ Block diagram illustration of the proposed trajectory planner for the reconnaissance phase (a). The method is composed of three major processes: a process to design firing timings (i.e., $t_1$ and $t_2$) and an attitude mode profile, a process to compute a desired delta-V and a process to assign desired attitudes to the nominal trajectory given by the previous processes. Figure (b) shows the overview of the process to design firing timings and an attitude mode profile.}
\label{fig:traplan}
\end{figure}

\section{Proposed Planner}~\label{sec:proposed_planner}
In this section, we present a proposed trajectory planner for the reconnaissance phase as shown in Figure~\ref{fig:missionconcept}, which involves (1) determining firing timings (i.e., $t_1$ and $t_2$), (2) designing desired mode profiles, and (3) solving a stochastic optimization problem.
We note that $t_1^o$ and $t_2^o$ are determined only by the position of the Sun and the sample site.
Thus, they are given before starting the planner (see Section~\ref{subsec:problem_formulation} for further details). At the end, we can design an attitude profile by assigning desired attitudes to each mode. Once the attitude mode profile and a state (\ie{}, position and velocity) profile are designed, desired attitudes are automatically assigned by using the attitude GNC subsystem in the~\ac{MuSCAT}.
Figure~\ref{fig:traplan} illustrates the overview of the proposed planner. 

\subsection{Pre-Designing Firing Timings and an Attitude Mode Profile}
Here, we detail the process to design firing timings and an attitude mode profile prior to solving a stochastic optimization problem. 
Specifically, there is an attitude mode selection history required to achieve the proposed reconnaissance phase, as shown in Figure~\ref{fig:attitude assign}.
In each mode, the attitude is precisely controlled to meet the specific requirements.
We note that each mode includes attitude change maneuvers to achieve a desired attitude at the beginning.
The green band represents a delta-V mode, the light blue band represents a sun-pointing mode, and the dark blue band represents the observation mode.
The delta-V mode involves changing an attitude to aim the chemical thruster in the desired direction and maintaining a certain attitude until the thruster is warmed to the required temperature.
In the sun-pointing mode, it controls the attitude to maximize power generation by aiming the solar panels toward the Sun.
Finally, in the observation mode, it controls the attitude to aim the camera at the bottom of the spacecraft to the sample site.
We note again that we take the nadir-relative pointing approach as the~\ac{OREX} mission did. Therefore, the desired attitude profile is computed only by the predicted position of the spacecraft given by the trajectory optimization process and the predicted position of the sample site.

Furthermore, we can create a reasonable attitude mode profile by considering the worst-case time required for a single attitude change maneuver to reach a steady state, the time required to heat the thrusters, and the time required to charge the battery SoC fully during the sun pointing mode.
We denote each these three times as $\Delta t_\text{attitude}$, $\Delta t_\text{heat}$, and $\Delta t_\text{SoC}$, respectively.
Delta-V modes are set to begin $\Delta t_\text{attitude}$ + $\Delta t_\text{heat}$ before the pre-scheduled firing timing.
Further, we set the beginning of observation mode $\Delta t_\text{attitude}$ before $t_1^o$.
It is important to note that each mode must last at least $\Delta t_\text{attitude}$ to have sufficient time for an attitude change maneuver.
Moreover, each instance of the solar pointing mode during $t_0$ and $t_3$ must last for at least $\Delta t_\text{attitude} + \Delta t_\text{SoC}$. Using the conceptual model, $\Delta t_\text{attitude}$ is provided as $30$ minutes whereas $\Delta t_\text{heat}$ is set to $30$ minutes and $\Delta t_\text{SoC}$ is set to $20$ minutes.
As for $\Delta t_\text{attitude}$, we measure the expected time to reach a steady state under the worst-case scenario, performing a flip maneuver in a system with actuator noise.
The $\Delta t_\text{heat}$ value is then set to the required preheating time at nominal for the chemical thruster assumed in the conceptual model.
Finally, using the same simulation environment in Section~\ref{subsec:decouple}, we measured the expected time to charge the battery from empty to full during the sun-pointing mode.
Given that the Time of Flight (TOF) for the reconnaissance phase is more than nine hours, we could reasonably schedule firing timings and modes satisfying all the above constraints before the computation, as in Figure~\ref{fig:attitude assign}.

\subsection{Problem Formulation for the Trajectory Optimization Process}~\label{subsec:problem_formulation}
Following the preceding discussion in Section~\ref{sec:mission_verification}, we introduce the stochastic optimization problem considered in the proposed trajectory planner, whose solution provides a nominal trajectory satisfying all of the mission constraints.
Setting $t_0$ to $\SI{0}{s}$, we begin with the discretized dynamics with uncertainties of the form: 
\begin{equation}
    x_{k} = 
    \begin{cases}
         F(x_{k-1}, \,\, k-1) + [0_{3\times1},\,\, \Delta v_k^\top]^{\top} + \beta [0_{3\times1}, \,\, (\Delta v_k \circ \omega_k)^\top]^{\top} \quad &(k\Delta t \in \{0, t_1, t_2, t_f\})\\
        F(x_{k-1}, \,\,  k-1) \quad &(k\Delta t \notin \{0, t_1, t_2, t_f\})
    \end{cases}, \label{eq:dynamicsconsts}
\end{equation}
where $x_k \in \mathbb{R}^{6}$ is the system state, $\Delta t \in \mathbb{R}$ is the discretized time step, and $\Delta v_k \in \mathbb{R}^{3}$ is the deterministic delta-V. Let $\Delta t = \frac{t_f}{N}$.
We note that the second term in Eq.~\eqref{eq:dynamicsconsts} represents the delta-V input modeled as an impulsive input in discretized time space, which was discussed in Section~\ref{subsec:model_assurance}.
The value $\beta \in \mathbb{R}$ is a coefficient parameter and $\omega_k \in \mathbb{R}^3$ is a normal Gaussian white noise representing control uncertainty.
Further, $F:\mathbb{R}^{6} \times \mathbb{R} \rightarrow \mathbb{R}^{6}$ is the discretized update equation of Eq.~\eqref{eq:doubleintegdyn}.

The magnitude of $\Delta v_k$ must be bounded between $0$ and the given constant parameter $\Delta v_\text{max} \in \mathbb{R}$ for all firings during the reconnaissance phase. This implies that
\begin{equation}
    \| \Delta v_k \|_2 \leq \Delta v_\text{max}, \quad k \Delta t \in \{ 0, t_1, t_2, t_f\}. \label{eq:controlconsts}
\end{equation}

In this work, we impose the boundary constraint to the expected value of the position.
Together with the fact that the state estimation has uncertainty distributed according to a Gaussian distribution, we have 
\begin{eqnarray}
    x_0 \sim \mathcal{N}(x_{t_0}, \Sigma_{t_0}), \quad \mathbb{E}(x_N) = x_{t_f},  \label{eq:boundaryconsts}
\end{eqnarray}
where $x_{t_0}$, $\Sigma_{t_0}$, and $x_{t_f}$ are given parameters. 
\subsubsection{Modeling State Constraints}
\label{modelstateconsts}

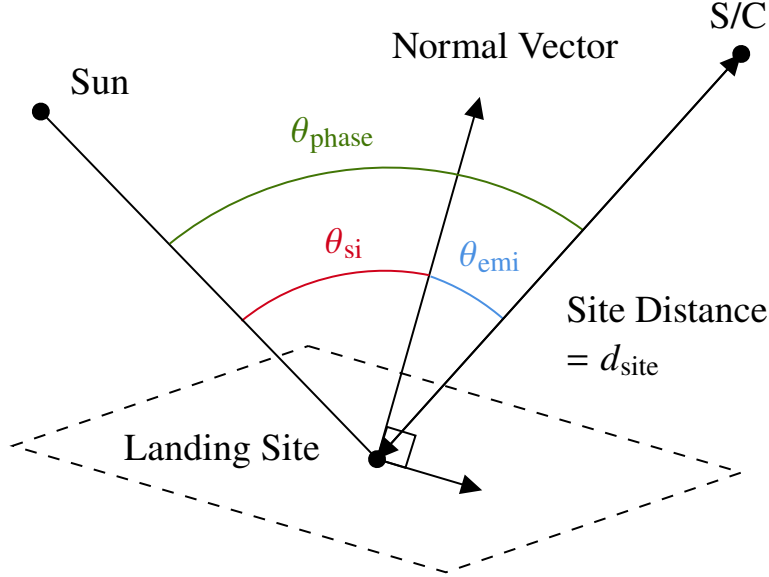
\begin{figure}[t!]
    \centering
    \tikzset{every picture/.style={line width=0.75pt}} 

\begin{tikzpicture}[x=0.75pt,y=0.75pt,yscale=-1,xscale=1, scale = 0.6]


\draw [line width=0.75]    (325.77,402.27) -- (411.29,427.65) ;
\draw [shift={(414.17,428.5)}, rotate = 196.53] [fill={rgb, 255:red, 0; green, 0; blue, 0 }  ][line width=0.08]  [draw opacity=0] (17.86,-8.58) -- (0,0) -- (17.86,8.58) -- cycle    ;
\draw    (411.35,100.39) -- (325.77,402.27) ;
\draw [shift={(412.17,97.5)}, rotate = 105.83] [fill={rgb, 255:red, 0; green, 0; blue, 0 }  ][line width=0.08]  [draw opacity=0] (17.86,-8.58) -- (0,0) -- (17.86,8.58) -- cycle    ;
\draw  [draw opacity=0] (211.87,286.01) .. controls (234.53,267.49) and (262.55,253.87) .. (294.13,247.56) .. controls (320.07,242.38) and (345.74,242.66) .. (369.68,247.57) -- (325.77,402.27) -- cycle ; \draw  [color={rgb, 255:red, 208; green, 2; blue, 27 }  ,draw opacity=1 ] (211.87,286.01) .. controls (234.53,267.49) and (262.55,253.87) .. (294.13,247.56) .. controls (320.07,242.38) and (345.74,242.66) .. (369.68,247.57) ;
\draw    (43.17,109.88) -- (325.77,402.27) ;
\draw [shift={(43.17,109.88)}, rotate = 45.97] [color={rgb, 255:red, 0; green, 0; blue, 0 }  ][fill={rgb, 255:red, 0; green, 0; blue, 0 }  ][line width=0.75]      (0, 0) circle [x radius= 6.7, y radius= 6.7]   ;
\draw  [draw opacity=0] (370.72,248.55) .. controls (390.75,255.63) and (410.09,266.35) .. (427.67,280.75) .. controls (429,281.84) and (430.32,282.94) .. (431.61,284.06) -- (325.77,402.27) -- cycle ; \draw  [color={rgb, 255:red, 74; green, 144; blue, 226 }  ,draw opacity=1 ] (370.72,248.55) .. controls (390.75,255.63) and (410.09,266.35) .. (427.67,280.75) .. controls (429,281.84) and (430.32,282.94) .. (431.61,284.06) ;
\draw  [draw opacity=0] (151.58,220.97) .. controls (186.21,192.96) and (228.82,172.37) .. (276.79,162.79) .. controls (359.18,146.32) and (439.76,165.83) .. (499.16,209.75) -- (325.77,402.27) -- cycle ; \draw  [color={rgb, 255:red, 65; green, 117; blue, 5 }  ,draw opacity=1 ] (151.58,220.97) .. controls (186.21,192.96) and (228.82,172.37) .. (276.79,162.79) .. controls (359.18,146.32) and (439.76,165.83) .. (499.16,209.75) ;
\draw   (334.26,375.15) -- (358.12,382.44) -- (349.63,409.56) -- (325.77,402.27) -- cycle ;
\draw    (629.9,63.65) -- (327.78,400.04) ;
\draw [shift={(325.77,402.27)}, rotate = 311.93] [fill={rgb, 255:red, 0; green, 0; blue, 0 }  ][line width=0.08]  [draw opacity=0] (17.86,-8.58) -- (0,0) -- (17.86,8.58) -- cycle    ;
\draw [shift={(631.9,61.42)}, rotate = 131.93] [fill={rgb, 255:red, 0; green, 0; blue, 0 }  ][line width=0.08]  [draw opacity=0] (17.86,-8.58) -- (0,0) -- (17.86,8.58) -- cycle    ;
\draw    (631.9,61.42) -- (325.77,402.27) ;
\draw [shift={(325.77,402.27)}, rotate = 131.93] [color={rgb, 255:red, 0; green, 0; blue, 0 }  ][fill={rgb, 255:red, 0; green, 0; blue, 0 }  ][line width=0.75]      (0, 0) circle [x radius= 6.7, y radius= 6.7]   ;
\draw [shift={(631.9,61.42)}, rotate = 131.93] [color={rgb, 255:red, 0; green, 0; blue, 0 }  ][fill={rgb, 255:red, 0; green, 0; blue, 0 }  ][line width=0.75]      (0, 0) circle [x radius= 6.7, y radius= 6.7]   ;
\draw  [dash pattern={on 4.5pt off 4.5pt}] (267.9,307.1) -- (632.06,413.51) -- (383.64,497.45) -- (19.48,391.04) -- cycle ;

\draw (298,220.38) node  [font=\Large,color={rgb, 255:red, 74; green, 144; blue, 226 }  ,opacity=1 ,rotate=-358.71]  {$\textcolor[rgb]{0.82,0.01,0.11}{\theta }\textcolor[rgb]{0.82,0.01,0.11}{_\textrm{si}}$};
\draw (420.58,231.54) node  [font=\Large,color={rgb, 255:red, 74; green, 144; blue, 226 }  ,opacity=1 ,rotate=-358.71]  {$\theta_\textrm{emi}$};
\draw (288.4,128.78) node  [font=\Large,color={rgb, 255:red, 74; green, 144; blue, 226 }  ,opacity=1 ,rotate=-358.71]  {$\textcolor[rgb]{0.25,0.46,0.02}{\theta }\textcolor[rgb]{0.25,0.46,0.02}{_\textrm{phase}}$};
\draw (65.41,70.59) node [anchor=north west][inner sep=0.75pt]  [font=\Large] [align=left] {Sun};
\draw (602.86,15.1) node [anchor=north west][inner sep=0.75pt]  [font=\Large] [align=left] {S/C};
\draw (335.96,40.56) node [anchor=north west][inner sep=0.75pt]  [font=\Large] [align=left] {Normal Vector};
\draw (482.4,262.91) node [anchor=north west][inner sep=0.75pt]  [font=\Large] [align=left] {Site Distance\\$\displaystyle =d_\textrm{site}$};
\draw (110.82, 378.85) node [anchor=north west][inner sep=0.75pt]  [font=\Large] [align=left] {Landing Site };

\end{tikzpicture}
    \caption{Graphical description of observation constraints.}
      \label{fig:angledif}
\end{figure}

\begin{table}[]
    \centering
    \begin{tabular}{ccccccc}
    \hline
          Site Distance & Bennu Local Solar Time & Emission Angle & Phase Angle & Solar Incidence Angle \\ 
          $d_\text{site}$ & $t_\text{LST}$ & $\theta_\text{emi}$ & $\theta_\text{phase}$ & 
          $\theta_\text{si}$ \\\hline
          $d_1 \leq d_\text{site} \leq d_2$ & 
          $t_{l1} \leq t_\text{LST}  \leq t_{l2}$ & 
          $\theta_{e1} \leq \theta_\text{emi}  \leq \theta_{e2} $&
          $\theta_{p1} \leq \theta_\text{phase}  \leq \theta_{p2} $&
          $\theta_{s1} \leq \theta_\text{si}   \leq  \theta_{s2}$  \\ \hline
    \end{tabular}
    \caption{Observation constraints for the reconnaissance phase}
    \label{tab:observabilityconsts}
\end{table}

\textit{Collision Avoidance Constraints: }
Here we focus on formulating state constraints during the reconnaissance phase.
We begin with safety-related constraints. To address safety concerns, the distance from the small body center of mass must be greater than the minimum threshold during the phase:
\begin{eqnarray}
    \|r_k \|_2 \geq h_\text{min}, \quad k = 0, ..., N, \label{eq:safetyconsts}
\end{eqnarray}
where $h_\text{min} \in \mathbb{R}$ is a parameter. 
We note that there exist additional safety constraints for collision avoidance that could be introduced for cases when a thruster misfires or if the spacecraft state enters a safe mode. While the proposed framework can easily accommodate such constraints, in this work, we focus solely on constraints of the form Eq.~\eqref{eq:safetyconsts}, aligning them with those considered in the~\ac{OREX} mission~\cite{LevineWibbenEtAl2022}.

\textit{Observation Constraints: }
We then turn our attention to formulating observation constraints, which are critical to the quality of scientific observations and consequently inform the trajectory design.
In order to maximize the success rate of observation while being robust against uncertainty, these constraints needed to be met as long as possible~\cite{LevineWibbenEtAl2022}.
In line with the observation constraints provided to the trajectory design team during multiple reconnaissance phases in the~\ac{OREX} mission, we address the following observation constraints as shown in Table~\ref{tab:observabilityconsts}.
The definitions of those constraints are depicted in Figure~\ref{fig:angledif}.
The solar incidence angle bounds the possible observation times. 
Similarly, the local solar time restricts the timing and duration of observations.
These two constraints do not depend on the position of the spacecraft.
Emission, phase angle, and site distance are all dependent on the spacecraft position.

\textit{Observation Time Window: }
Before integrating these observation constraints into the trajectory optimization problem, it is necessary and possible to compute the maximum observation time window for each sample site, which is independent of the spacecraft pose.
It should be noted that those windows are defined only by the Solar Incidence Angle, Local Solar Time constraint, the relative position of the asteroid to the Sun, rotational period, sample site position, and the normal vector of the site surface.
Since all of these are independent of the spacecraft position, the boundaries of the time windows are given as constants before running the trajectory planner.
Throughout the rest of the paper, these predefined observation time windows are denoted as $t^{o}_1$ and $t^{o}_2$.
Next, we now model observation constraints in Table~\ref{tab:observabilityconsts}.
Site distance constraints, emission angle constraints, and phase angle constraints are defined as
\begin{eqnarray}
    d_1 \leq \| r_k - r_{\text{site},k}\| \leq d_2, \quad t^o_1 \leq k \Delta t \leq t^o_2, \label{eq:sitedistanceconsts} \\
   \cos \theta_{e2} \leq  \frac{(r_k - r_{\text{site},k})^\top r_{s\perp, k}}{\|r_k - r_{\text{site},k}\|_2} \leq \cos\theta_{e1}, \quad t^o_1 \leq k \Delta t \leq t^o_2, \label{eq:emissionangleconsts}  \\
   \cos\theta_{p2} \leq  \frac{(r_k - r_{\text{site},k})^\top \hat{r}_\text{center$\rightarrow$sun}}{\|r_k - r_{\text{site},k}\|_2} \leq \cos\theta_{p1}, \quad t^o_1 \leq k \Delta t \leq t^o_2, \label{eq:phaseangleconsts}
\end{eqnarray}
where $r_{\text{site},k}$ is the position of the sample site, $r_{s\perp, k}$ is the normal vector of the site surface, and $\hat{r}_\text{center$\rightarrow$sun}$ is the unit vector from the small body center of mass to the Sun. We note that the vector from the sample site to the Sun is approximated by $\hat{r}_\text{center$\rightarrow$sun}$ because the Bennu orbit distance from the Sun is the order of $100$ million kilometers, while the diameter of Bennu is the order of $0.5$ kilometers.

\textit{Battery Constraints: }
Following the discussion in Section~\ref{subsec:decouple}, we impose constraints such that the battery~\ac{SoC} is above the minimum threshold during the reconnaissance phase.
We assume that the spacecraft flies over the daytime side of the small body while observing the sample site on the surface of the small body.
Therefore, when considering the battery~\ac{SoC} during the observation, we impose a constraint such that the spacecraft is placed outside the cone on the sunlit side, which gives
\begin{gather}
   \frac{(\Delta v_k)^\top \hat{r}_\text{center$\rightarrow$sun}}{\| \Delta v_k\|_2} \leq \cos\theta_\text{deltaV}, \quad  
   \frac{(\Delta v_k)^\top (-\hat{r}_\text{center$\rightarrow$sun})}{\| \Delta v_k\|_2} \leq \cos\theta_\text{deltaV}, 
   \quad k \Delta t \in  \{0, t_1, t_2, t_f\}
   \label{eq:batterydeltavconsts}\\
   \frac{(r_k  - r_{\text{site},k})^\top \hat{r}_\text{center$\rightarrow$sun}}{\|r_k - r_{\text{site},k}\|_2} \leq  \cos\theta_\text{obs}, \quad t^o_1 \leq k \Delta t \leq t^o_2, \label{eq:batteryobsconsts}    
\end{gather}
where we apply the same approximation as in Eq.~\eqref{eq:phaseangleconsts}.

\textit{Chance Constraints Reformulation: }
Due to uncertainty in the dynamics, each state constraint must be satisfied with at least a probability of $p_\text{chance}$, which is a user-defined risk bound.
Therefore, all state constraints are reformulated in the form of chance constraints, which are given by
\begin{gather}
    \text{Pr}(\|r_k\|_2 \geq h_\text{min}) \geq p_\text{chance}, \quad k = 0, ..., N, \label{eq:safetychanceconsts} \\
 \text{Pr}\left(\frac{(r_k - r_{\text{site},k})^\top \hat{r}_\text{center$\rightarrow$sun}}{\|r_k - r_{\text{site},k}\|_2} \leq  \cos\theta_\text{obs} \right) \geq p_\text{chance}, \quad t^o_1 \leq k \Delta t \leq t^o_2, \label{eq:batteryobschanceconsts} \\
\text{Pr}(d_1 \leq \| r_k - r_{\text{site},k}\|_2 \leq d_2) \geq p_\text{chance}, \quad t^o_1 \leq k \Delta t \leq t^o_2, \label{eq:sitedistancechanceconsts} \\
   \text{Pr}\left(\cos\theta_{e2} \leq  \frac{(r_k - r_{\text{site},k})^\top r_{s\perp, k}}{\|r_k - r_{\text{site},k}\|_2} \leq \cos\theta_{e1})\right)\geq p_\text{chance}, \quad t^o_1 \leq k \Delta t \leq t^o_2, \label{eq:emissionanglechanceconsts} \\
   \text{Pr}\left(\cos\theta_{p2} \leq  \frac{(r_k - r_{\text{site},k})^\top \hat{r}_\text{center$\rightarrow$sun}}{\|r_k - r_{\text{site},k}\|_2} \leq \cos\theta_{p1}\right) \geq p_\text{chance}, \quad t^o_1 \leq k \Delta t \leq t^o_2. \label{eq:phaseanglechanceconsts}
\end{gather}
Finally, we formally define the stochastic minimum fuel optimization problem as follows:

\vspace{0.25cm}
\noindent\fbox{%
\parbox{0.98\linewidth}{%
\vspace{0.5em}
\centering{\text{\underline{Problem 1: Baseline stochastic optimization problem}}}
\begin{equation}
\tag{\textbf{P1}}
\begin{aligned}
      \min_{\Delta v_k} & \sum_{ k\Delta t \in \{ 0, t_1, t_2, t_f\}}\| \Delta v_k \|_2, \\
    \text{ s.t. } 
    &\quad \text{Eq.~\eqref{eq:dynamicsconsts},~\eqref{eq:controlconsts},~
    \eqref{eq:boundaryconsts},~
    \eqref{eq:batterydeltavconsts},~\eqref{eq:safetychanceconsts},~\eqref{eq:batteryobschanceconsts},~\eqref{eq:sitedistancechanceconsts},~\eqref{eq:emissionanglechanceconsts}, and~\eqref{eq:phaseanglechanceconsts} \thinspace.}     
 \end{aligned} \label{pro:problem1}
\end{equation}}} 
\vspace{0.25cm}

\section{A Solution Framework for Trajectory Optimization Process}~\label{sec:solution_algo}
This section details a proposed approach to solve~\ref{pro:problem1}.
By representing a state distribution as a Gaussian to capture uncertainty, we are able to introduce an~\ac{EKF}-inspired method for modeling stochastic dynamics as deterministic nonlinear dynamics.
The key difficulties in solving~\ref{pro:problem1} stem from the presence of non-convex chance constraints and the lack of a closed form expression for satisfying them.
Such difficulties motivated an approach to approximate them as the intersection of affine chance-constraints.
In order to accomplish this, we propose a solution framework that entails 1) solving an optimization problem without uncertainties, 2) propagating through the modeled deterministic nonlinear dynamics to complete the approximation, and 3) solving an optimization problem considering uncertainties with the solution from the first two steps as an initial guess for a solver. 

\subsection{Gaussian Representation to Capture Uncertainty}
In order to solve an optimal control problem with these system dynamics, we must reformulate the stochastic dynamics as deterministic functions.
\textcolor{black}{Provided that Section~\ref{subsec:model_uncertainty} has demonstrated that the Gaussian assumption is reasonable to capture uncertainties propagated through nonlinear dynamics, we approximate the probability distribution of $x_k$ as a Gaussian distribution.}
Further, we employ the prediction step of~\ac{EKF} to model uncertainty propagation.
When the Gaussian distribution of the state at the beginning of the phase is known, \ie{}, the mean and covariance are given as $\mu_0 \in \mathbb{R}^{3}$ and $\Sigma_0\in \mathbb{R}^{3\times3}$, the mean and covariance dynamics are modeled as follows:
\begin{eqnarray}
    \mu_{k} &=& 
    \begin{cases}
         F(\mu_{k-1}, k-1) + [0_{3\times1}, \,\, \Delta v_k^\top]^\top \quad &(k\Delta t \in \{0, t_1, t_2, t_f\}),\\
         F(\mu_{k-1}, k-1) \quad &(k\Delta t \notin \{0, t_1, t_2, t_f\}),
    \end{cases} \label{eq:meandynamics} \\
    \Sigma_{k} &=& 
    \begin{cases}
         \nabla_x F \Sigma_{k-1} \nabla_x F^\top + [0_{3\times6};\,\,  0_{3\times3}, \,\, \beta v_k \circ I] I [0_{3\times6};\,\, 0_{3\times3}, \,\, \beta v_k \circ I]^\top  \quad &(k\Delta t \in \{0, t_1, t_2, t_f\}),\\
         \nabla_x F \Sigma_{k-1} \nabla_x F^\top \quad &(k\Delta t \notin \{0, t_1, t_2, t_f\}),
    \end{cases} \label{eq:covdynamics}
\end{eqnarray}
where $\nabla_x F$ is the Jacobian matrix of $F$ with respect to $x_k$ and it is evaluated at $(x_k, k) = (\mu_{k-1}, k-1)$. 
Given that $\Sigma_k$ and $\mu_k$ are regarded as decision variables, boundary constraints are expressed as linear constraints as
\begin{gather}
    \mu_o = x_{t_0}, \quad \mu_N = x_{t_f}, \label{eq:meanbound} \\
    \Sigma_0 = \Sigma_{t_0}.     \label{eq:covbound} 
\end{gather}

\subsection{Solution Approach}
Here, we focus on transforming the chance constraints into deterministic nonlinear constraints.
We begin by rewriting the chance constraints in ~\ref{pro:problem1} into constraints using a general form. 
Feasible regions of all events for the chance constraints in~\ref{pro:problem1} are classified into three types of sets: regions inside of a second-order cone, regions outside of a second-order cone, and regions being in the intersection of them.
The general form of a second order cone is expressed as 
\begin{eqnarray}
 SOC_i := \{x \in \mathbb{R}^{6}\,\, | \,\, \| A_i x + b_i \|_2 \leq C_i x + d_i \}, \label{eq:SOC_def}
\end{eqnarray}
where $i$ denotes indices of constraints, and $A_i \in \mathbb{R}^{6\times6}$, $b_i \in \mathbb{R}^{6}$, $C_i \in \mathbb{R}^{6\times6}$, and $d_i \in \mathbb{R}^{6}$ are parameters specified for each.
Subsequently, all of the chance constraints Eq.\eqref{eq:safetychanceconsts},~\eqref{eq:batteryobschanceconsts},~\eqref{eq:sitedistancechanceconsts},~\eqref{eq:emissionanglechanceconsts}, and~\eqref{eq:phaseanglechanceconsts} are rewritten as
\begin{gather}
    \text{Eq.}\eqref{eq:safetychanceconsts} \Leftrightarrow \text{Pr}(x \notin SOC_1) \geq p_\text{chance}, \quad k = 0, ..., N, \label{eq:safetychanceconstsapp1}  \\
    \text{Eq.}\eqref{eq:batteryobschanceconsts} \Leftrightarrow \text{Pr}(x \notin SOC_2) \geq p_\text{chance}, \quad t^o_1 \leq k \Delta t \leq t^o_2,  \label{eq:batteryobschanceconstsapp1}  \\ 
    \text{Eq.}\eqref{eq:sitedistancechanceconsts} \Leftrightarrow \text{Pr}(x \notin SOC_3 \cap x \in SOC_4 ) \geq p_\text{chance},\quad t^o_1 \leq k \Delta t \leq t^o_2, \label{eq:sitedistancechanceconstsapp1} \\
    \text{Eq.}\eqref{eq:emissionanglechanceconsts}\Leftrightarrow \text{Pr}(x \notin SOC_5 \cap x \in SOC_6 ) \geq p_\text{chance},\quad t^o_1 \leq k \Delta t \leq t^o_2, \label{eq:emissionanglechanceconstsapp1}\\
    \text{Eq.}\eqref{eq:phaseanglechanceconsts} \Leftrightarrow \text{Pr}(x \notin SOC_7 \cap x \in SOC_8 ) \geq p_\text{chance}, \quad t^o_1 \leq k \Delta t \leq t^o_2.  \label{eq:phaseanglechanceconstsapp1}
\end{gather}

Then, the joint chance constraints presented here boil down to the conjunction of two chance constraints.
A common approach to relax a joint chance constraint is to use Boole's inequality, which replaces it with two chance constraints given by
\begin{gather}
    \text{Pr}(x \notin SOC_i) \cap x \in SOC_j) \geq 1 - (1 - p_\text{chance}), \\
    \sim \text{Pr}(x \notin SOC_i)  \geq 1 - \frac{1 - p_\text{chance}}{2}, \quad \text{Pr}(x \in SOC_j) \geq 1 - \frac{1 - p_\text{chance}}{2}.
\end{gather}

\begin{figure}[t!]
     \centering
     \begin{subfigure}[b]{0.45\textwidth}
         \centering
         \includegraphics[width=\textwidth]{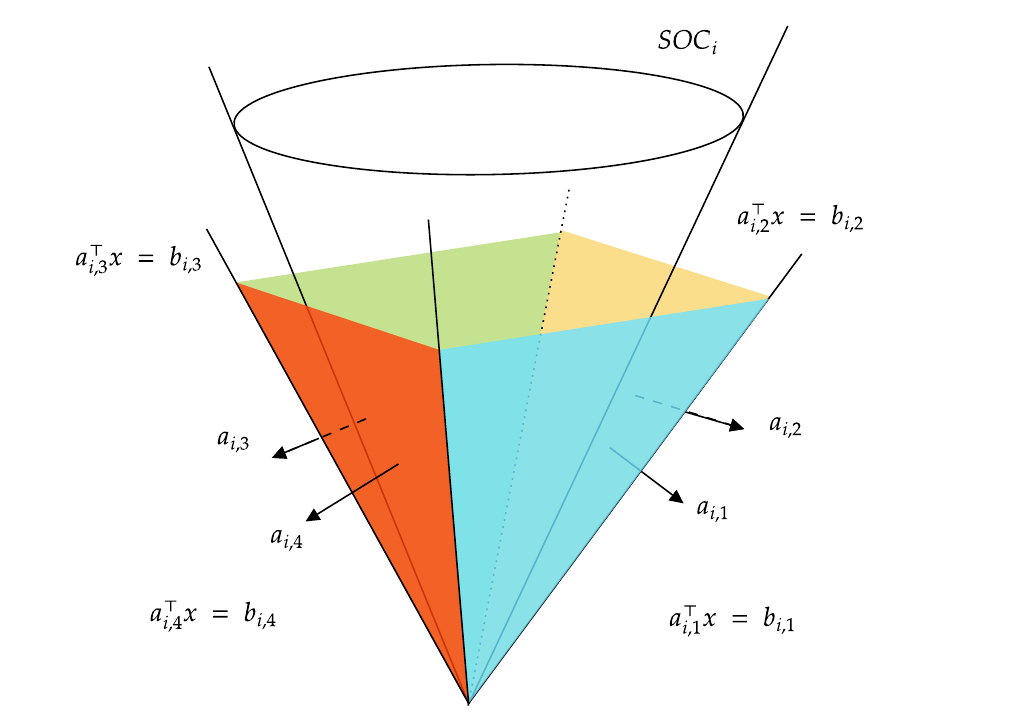}
         \caption{Approximation of $x \in \text{SOC}_i$ by the conjunction of affine constraints.}
     \end{subfigure}
     \hfill
     \begin{subfigure}[b]{0.45\textwidth}
         \centering
         \includegraphics[width=\textwidth]{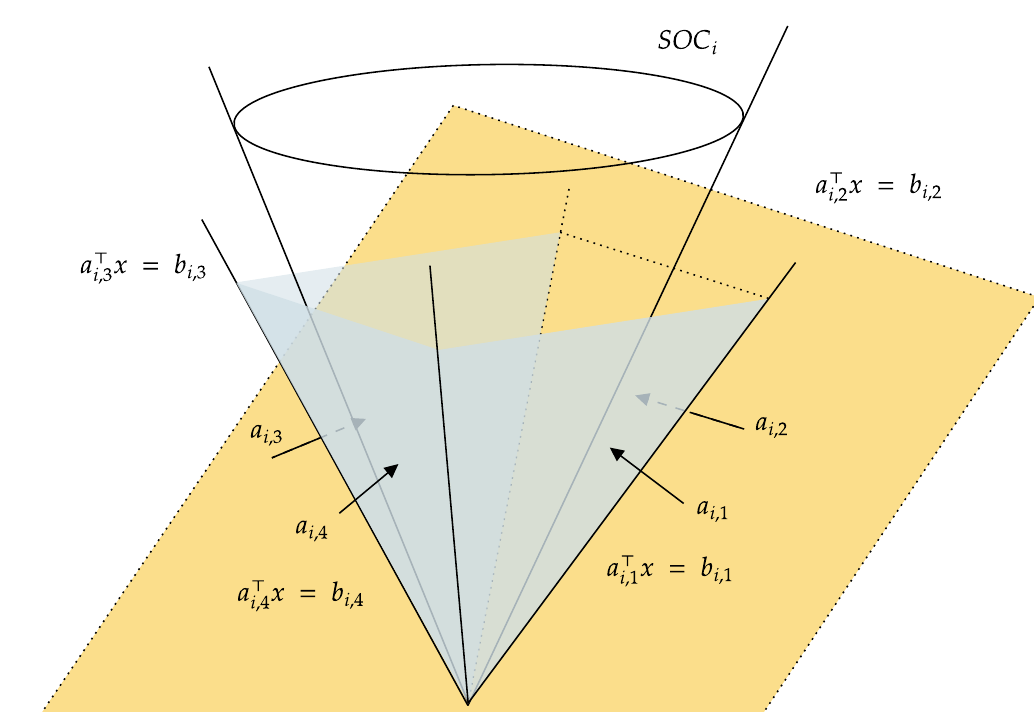}
         \caption{Approximation of $x \notin \text{SOC}_i$ by the disjunction of affine constraints. We choose a hyperplane such that it maximizes $\text{Pr}(x \in A_{i,j})$. The other hyperplanes are deactivated (colored in gray).}
     \end{subfigure}
        \caption{Polytopic approximation of a second order cone set $SOC_i$. The parameters of hyperplanes can be given before the computation. Choosing the number of hyperplanes can approximate the second order cone with arbitrary precision.}
        \label{fig:conceptSOCP}
\end{figure}

In order to address those two chance constraints, we consider the polytopic approximation of this second order cone, as seen in Figure~\ref{fig:conceptSOCP}.
Its key advantage is, with the Gaussian approximation, affine chance constraints can be rigorously transformed into determistic ones.
We note that $a_{i,j} \in \mathbb{R}^6$ and $b_{i,j} \in \mathbb{R}^6$ forming each hyperplane can be computed beforehand and we can choose an arbitrary number of hyperplanes in the need of approximation accuracy.
As a result, $x\in \text{SOC}_i$ is transformed into the conjunction of finitely many affine constraints whereas $x\notin \text{SOC}_i$ is transformed into the disjunction of them. Denoting the set bounded by a hyperplane as $A_{i,j} := \{x \,\, |\,\, a_{i,j}^\top x \leq b_{i,j} \}$, we have
\begin{eqnarray}
    \text{Pr}(x \notin SOC_i) \sim  \text{Pr}\left( \bigcup\limits_{j = 1}^{n_i} x \in A_{i,j}\right), \quad 
    \text{Pr}(x \in SOC_i) \sim  \text{Pr}\left( \bigcap\limits_{j = 1}^{n_i} x \in A_{i,j}\right).
\end{eqnarray}

We can now transform those two chance constraints into deterministic ones.
Using the typical approach with Boole's inequality again, we have
\begin{eqnarray}
    \text{Pr}\left( \bigcap\limits_{j = 1}^{n_i} x \in A_{i,j}\right) \geq 1 - (1 - p_i) 
        &\sim& \text{Pr}(x \in A_{i,j})  \geq 1 - \frac{1 - p_i}{n_i}, \,\, j = 1,...,n_i, \\
        &\Leftrightarrow& a_{i,j}^\top \mu_k + \Psi^{-1}\left(1 - \frac{(1 - p_i)}{n_i}\right) \sqrt{a_{i,j}^{\top} \Sigma_k a_{i,j}} \leq b_{i,j}, \quad j = 1,...,n_i, \\
    \text{Pr}\left( \bigcup\limits_{j = 1}^{n_i} x \in A_{i,j} \right) \geq 1 - (1 - p_i) 
        &\Leftarrow& \max_{j = 1,...,n_i} \text{Pr}(x \in A_{i,j}) \geq 1 - (1 - p_i)  \\
        &\Leftrightarrow& a_{i,\hat{j}}^\top \mu_k + \Psi^{-1}(1 - (1 - p_i)) \sqrt{a_{i,\hat{j}}^{\top} \Sigma_k a_{i,\hat{j}}} \leq b_{i,\hat{j}}, 
\end{eqnarray}
where $\hat{j} = \argmax_{j = 1,..., n_i} \text{Pr}(x \in A_{i,j})$ and $p_i \in \mathbb{R}$ represents a probability for each constraint after relaxing joint chance constraints.
We note that $\Psi^{-1}(\alpha)$ is the standard normal quantile and it can be calculated before the computation.
Despite the fact that there are several existing approaches to provide more accurate approximation, we have found that the proposed framework provides a feasible solution successively in practice.
Finally, all of the chance constraints are approximated as
\begin{gather}
    \text{Eq.}\eqref{eq:safetychanceconstsapp1} \sim a_{1,\hat{j}}^\top \mu_k + \Psi^{-1}(1 - (1 - p_1)) \sqrt{a_{1,\hat{j}}^{\top} \Sigma_k a_{1,\hat{j}}} \leq b_{1, \hat{j}}, \quad k = 0, ..., N. \label{eq:safetychanceconstsapp2} \\
    \text{Eq.}\eqref{eq:batteryobschanceconstsapp1} \sim a_{2,\hat{j}}^\top \mu_k + \Psi^{-1}(1 - (1 - p_2)) \sqrt{a_{2,\hat{j}}^{\top} \Sigma_k a_{2,\hat{j}}} \leq b_{2,\hat{j}}, \quad t^o_1 \leq k \Delta t \leq t^o_2  \label{eq:batteryobschanceconstsapp2}  \\ 
    \text{Eq.}\eqref{eq:sitedistancechanceconstsapp1} \sim 
    \begin{cases}
    a_{3,\hat{j}}^\top \mu_k + \Psi^{-1}(1 - (1 - p_3)) \sqrt{a_{3,\hat{j}}^{\top} \Sigma_k a_{3,\hat{j}}} \leq b_{3,\hat{j}}, \\
    a_{4,j}^\top \mu_k + \Psi^{-1}\left(1 - \frac{(1 - p_4)}{n_4}\right) \sqrt{a_{4,j}^{\top} \Sigma_k a_{4,j}} \leq b_{4,j}, \quad j = 1,...,n_4,         
    \end{cases} \quad t^o_1 \leq k \Delta t \leq t^o_2 \label{eq:sitedistancechanceconstsapp2} \\
\text{Eq.}\eqref{eq:emissionanglechanceconstsapp1} \sim 
    \begin{cases}
    a_{5,\hat{j}}^\top \mu_k + \Psi^{-1}(1 - (1 - p_5)) \sqrt{a_{5,\hat{j}}^{\top} \Sigma_k a_{5,\hat{j}}} \leq b_{5,\hat{j}},  \\
    a_{6,j}^\top \mu_k + \Psi^{-1}\left(1 - \frac{(1 - p_6)}{n_6}\right) \sqrt{a_{6,j}^{\top} \Sigma_k a_{6,j}} \leq b_{6,j}, \quad j = 1,...,n_6,         
    \end{cases} \quad t^o_1 \leq k \Delta t \leq t^o_2 \label{eq:emissionanglechanceconstsapp2} \\    
\text{Eq.}\eqref{eq:phaseanglechanceconstsapp1} \sim 
    \begin{cases}
    a_{7,\hat{j}}^\top \mu_k + \Psi^{-1}(1 - (1 - p_7)) \sqrt{a_{7,\hat{j}}^{\top} \Sigma_k a_{7,\hat{j}}} \leq b_{7,\hat{j}},  \\
    a_{8,j}^\top \mu_k + \Psi^{-1}\left(1 - \frac{(1 - p_8)}{n_8}\right) \sqrt{a_{8,j}^{\top} \Sigma_k a_{8,j}} \leq b_{8,j}, \quad j = 1,...,n_8,         
    \end{cases} \quad t^o_1 \leq k \Delta t \leq t^o_2 \label{eq:phaseanglechanceconstsapp2}  
\end{gather}

\normalem

\begin{table}[t!]
        \centering
        \begin{tabular}{cc}
        \hline
        Parameter  & Value \\ \hline
        $d_1$, \, $d_2$  & $\SI{500}{m}$, \, $\SI{740}{m}$  \\
        $t_{l1}$, \, $t_{l2}$& $\SI{8}{am}$$^{\mathrm{*}}$, \, $\SI{4}{pm}$$^{\mathrm{*}}$ \\
        $\theta_{e1}$, \, $\theta_{e2}$ & $\SI{20}{\degree}$, \, $\SI{45}{\degree}$ \\
        $\theta_{p1}$, \, $\theta_{p2}$ & $\SI{10}{\degree}$, \, N/A \\
        $\theta_{s1}$, \, $\theta_{s2}$ & N/A, \, $\SI{60}{\degree}$ \\
        $\Delta v_\text{max}$ & $\SI{0.2}{m/s}$ \\
        $h_\text{min}$ & $\SI{500}{m}$ \\
        $\hat{r}_\text{center$\rightarrow$sun}$ & $[-0.9992, 0.0323, 0.0225]^\top\SI{}{m}$ \\
        $p_\text{chance}$ & $\SI{70}{\%}$ \\
        $\beta$ & $\SI{8e-2}{}$ \\
        $x_0$ & $[\SI{0.0702}{km}, \SI{-1.0963}{km}, \SI{0.4064}{km}$, $\SI{-0.0047}{m/s}, \SI{-0.0167}{m/s}, \SI{0.0278}{m/s}]^\top$ \\
        $x_f$  & $[\SI{0.0199}{km}, \SI{1.1287}{km}, \SI{-0.1409}{km}$, $\SI{0.0003}{m/s}, \SI{0.0496}{m/s}, \SI{-0.0457}{m/s}]^\top$ \\
        $\Sigma_{t_0}$ & $\SI{1e-4}{} * \text{diag}(x_0 \circ x_0)$ \\
        $t_0$, $t_1$, $t_2$, $t_f$ & $\SI{0}{s}$, $\SI{8000}{s}$, $\SI{25000}{s}$, $\SI{34842}{s}$ \\
        $t_1^o$, $t_2^o$ & $\SI{13140}{s}$, $\SI{16537}{s}$ \\ 
        $r_{s \perp, k}$ & $T_{1,k}[0.394316, 0.565411, 0.724448]^{\top{\mathrm{**}}}$ \\ 
        $\Delta t$ & \SI{232}{s} \\\hline
        \multicolumn{2}{l}{$^{\mathrm{*}}$ Bennu Local Solar Time} \\
        \multicolumn{2}{l}{$^{\mathrm{**}}$ $T_{1,k}$ is a transformation matrix from the~\ac{BBF} to the J2000 frame.} \\
       \end{tabular}
     \caption{Numerical parameters for the OSIRIS-REx Reconnaissance B trajectory optimization using the proposed planner.}
     \label{tab:parameters}
\end{table}

Consequently, we could transform the stochastic optimization problem~\ref{pro:problem1} into the following deterministic optimization problem.
The constraints in this problem are all convex without Eq.~\eqref{eq:meandynamics}, Eq.~\eqref{eq:covdynamics}, and Eq.~\eqref{eq:batterydeltavconsts}, allowing the proposed formulation to be solved by the off-the shelf nonlinear optimization solver such as~\texttt{ipopt}.

\vspace{0.25cm}
\noindent\fbox{%
\parbox{0.98\linewidth}{%
\vspace{0.5em}
\centering{\text{\underline{Problem 2: Approximated deterministic optimization problem}}}
\begin{equation}
\tag{\textbf{P2}}
\begin{aligned}
      \min_{\Delta v_k} & \sum_{ k\Delta t \in \{ 0, t_1, t_2, t_f\}}\| \Delta v_k \|_2, \\
    \text{ s.t. } 
    &\quad  \text{Eq.~\eqref{eq:meandynamics},~\eqref{eq:covdynamics},~\eqref{eq:controlconsts},~\eqref{eq:meanbound},~\eqref{eq:covbound},~\eqref{eq:batterydeltavconsts},~\eqref{eq:safetychanceconstsapp2},~\eqref{eq:batteryobschanceconstsapp2},~\eqref{eq:sitedistancechanceconstsapp2},~\eqref{eq:emissionanglechanceconstsapp2}, and~\eqref{eq:phaseanglechanceconstsapp2} \, .}   
 \end{aligned} \label{pro:problem2}
\end{equation}}}
\vspace{0.25cm}

The next immediate question is how to detect $\hat{j}$ maximizing $\text{Pr}(x \in A_{i,j})$.
In order to address it, we propose a framework consisting of 1) solving the optimization problem without uncertainty as in~\ref{pro:problem3}, 2) propagating covariance throughout the modeled dynamics as in Eq.~\eqref{eq:meandynamics} and Eq.~\eqref{eq:covdynamics} to detect $\hat{j}$ for each constraint, and 3) solving the deterministic optimization problem considering uncertainty as in~\ref{pro:problem2}. 
Given that the current off-the-shelf solver for nonlinear optimization problems requires an initial guess~\cite{WachterBiegler2006}, which significantly impacts solution quality, this multi-stage approach offers the advantage of enhanced numerical stability. 

\vspace{0.25cm}
\noindent\fbox{%
\parbox{0.98\linewidth}{%
\vspace{0.5em}
\centering{\text{\underline{Problem 3: Deterministic optimization problem without uncertainty}}}
\begin{equation}
\tag{\textbf{P3}}
\begin{aligned}
      \min_{\Delta v_k} & \sum_{ k\Delta t \in \{ 0, t_1, t_2, t_f\}}\| \Delta v_k \|_2, \\
    \text{ s.t. } 
    &\quad  \text{Eq.~\eqref{eq:meandynamics}, ~\eqref{eq:controlconsts},~\eqref{eq:meanbound},~\eqref{eq:batterydeltavconsts},~\eqref{eq:safetyconsts},~\eqref{eq:batteryobsconsts},~\eqref{eq:sitedistanceconsts},~\eqref{eq:emissionangleconsts}, and~\eqref{eq:phaseangleconsts} \, .}   
 \end{aligned} \label{pro:problem3}
\end{equation}}}
\vspace{0.25cm}

\section{Numerical Examples}~\label{sec:numerical_examples}


\begin{figure}[t!]
    \centering
    \includegraphics[width=0.8\textwidth]{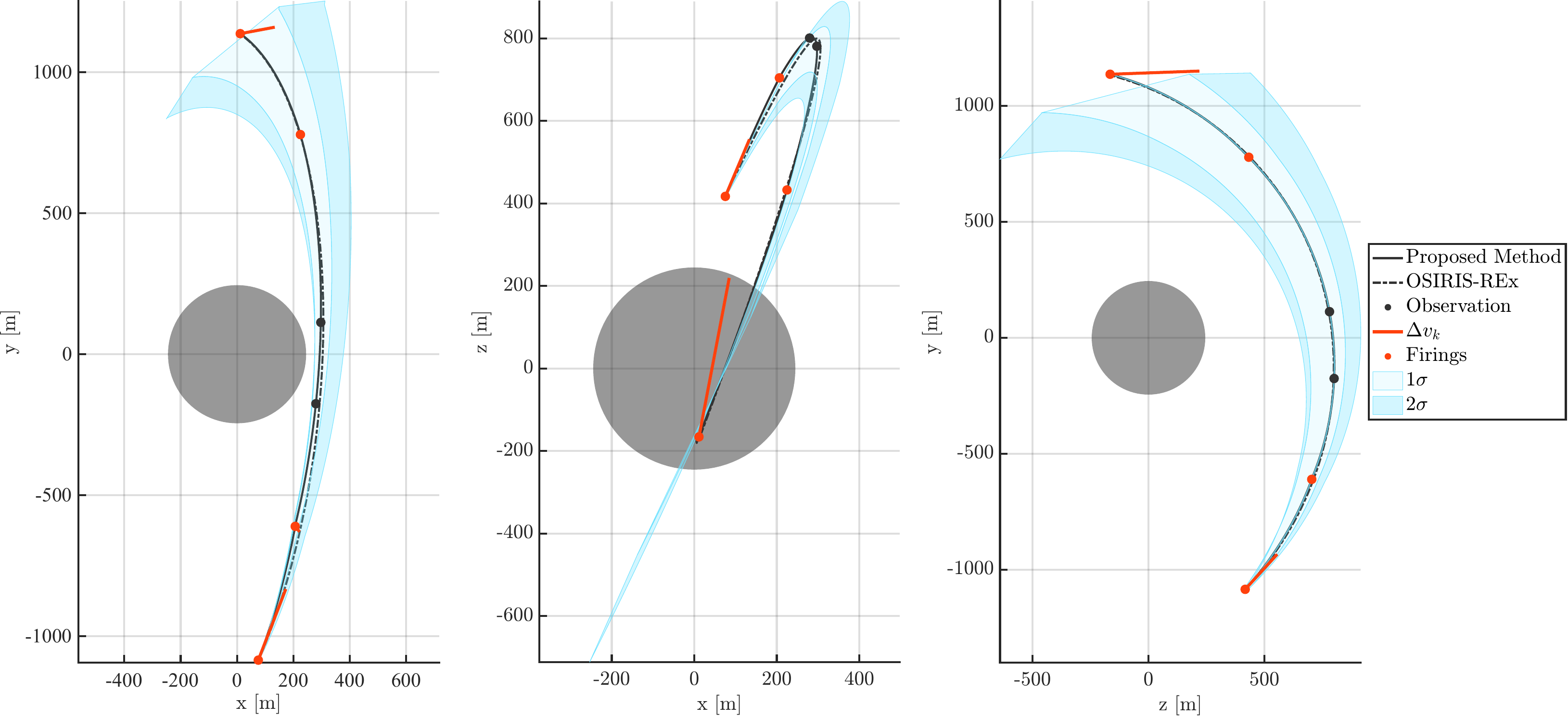}
    \caption{Projected views of the converged trajectory and the as-flown OSIRIS-REx trajectory in the J2000 frame (black line and black dotted line). Projected figures also display the circular approximated shape of Bennu (gray sphere), the delta-V firings (red vectors), the states at the timing of firings or the boundaries of the observation (red and black circular markers), and the $1\sigma$ and $2\sigma$ dispersions (light blue and blue highlighted areas). Note that the red vectors representing the delta-V firings are scaled by size, such that only their relative magnitudes are meaningful. Given that the worst $2\sigma$ dispersion does not collide with Bennu, we conclude that the converged solution satisfies the safety constraint as in Eq.~\eqref{eq:safetychanceconsts}.}
    \label{fig:result_1}
\end{figure}

This section presents implementation details and some numerical results to demonstrate the capability of the proposed approach. Specifically, the proposed approach is tested in the Reconnaissance B scenario for the Nightingale and compared against the as-flown trajectory executed for the~\ac{OREX}~\cite{LevineWibbenEtAl2022,WilliamsAntreasianEtAl2018}.
The simulations are implemented in MATLAB using the~\texttt{CasADi} interface~\cite{AnderssonGillisEtAl2019} and~\texttt{ipopt}~\cite{WachterBiegler2006} as a nonlinear programming solver. 
In order to get the as-flown trajectory and model nonlinear dynamics, the trajectory information and gravitational information from the~\ac{OREX} SPICE Kernels are used~\cite{ACTON199665,ACTON20189}.
Table~\ref{tab:parameters} summarizes all parameters used for this simulation.
For the proposed planner, we define $a_{i,j}$ and $b_{i,j}$ as follows: first, constraints whose boundaries are spheres or ellipsoids (\ie{}, $C_i$ in Eq.~\eqref{eq:SOC_def} equals zero) are approximated as the affine transformations of the following polytope:
\begin{gather}
P = \text{conv}(V), \quad \text{where} \\
V = \left\{[\pm1, 0, 0]^\top, [0, \pm1, 0]^\top, [0, 0, \pm 1]^\top, [\pm \sqrt{\frac{1}{3}}, \pm \sqrt{\frac{1}{3}}, \pm \sqrt{\frac{1}{3}}]^\top\right\}.
\end{gather}
Next, constraints whose boundaries are cones (i.e., $C_i$ in Eq.~\eqref{eq:SOC_def} is not zero) are approximated as the affine transformations of the following constraint:
\begin{equation}
    \begin{bmatrix}
            \cos\theta & 0 & -\sin \theta \\
            -\cos\theta & 0 & -\sin \theta \\
            0              & \cos\theta &  -\sin \theta \\
            0              & -\cos\theta &  -\sin \theta \\
    \end{bmatrix} x \leq 0, 
\end{equation}
where it over-approximates the following conic constraint:
\begin{equation}
    [0, 0, 1] x \geq \|x\|_2 \cos\theta.
\end{equation}
Finally, for the constraints $x \in SOC$, those approximations are affinely transformed such that the approximated hyperplanes under-approximate the feasible sets of the original constraints. 
Meanwhile, for the constraints $x \notin SOC$, they are affinely transformed such that the approximated hyperplanes over-approximate them. 
The proposed planner also requires an initial trajectory guess, for which the as-flown trajectory is chosen.
It is referred to as a ``warm-start.'' We note that we will also address a ``cold-start'' where a linear interpolation from $x_0$ to $x_f$ is chosen as an initial guess with a constant control vector $\Delta v_k = [1e^{-10}, 1e^{-10}, 1e^{-10}]^\top$ for every firing.

\begin{figure}[t!]
     \centering
     \begin{subfigure}[b]{0.25\textwidth}
         \centering
    \includegraphics[width=\textwidth]{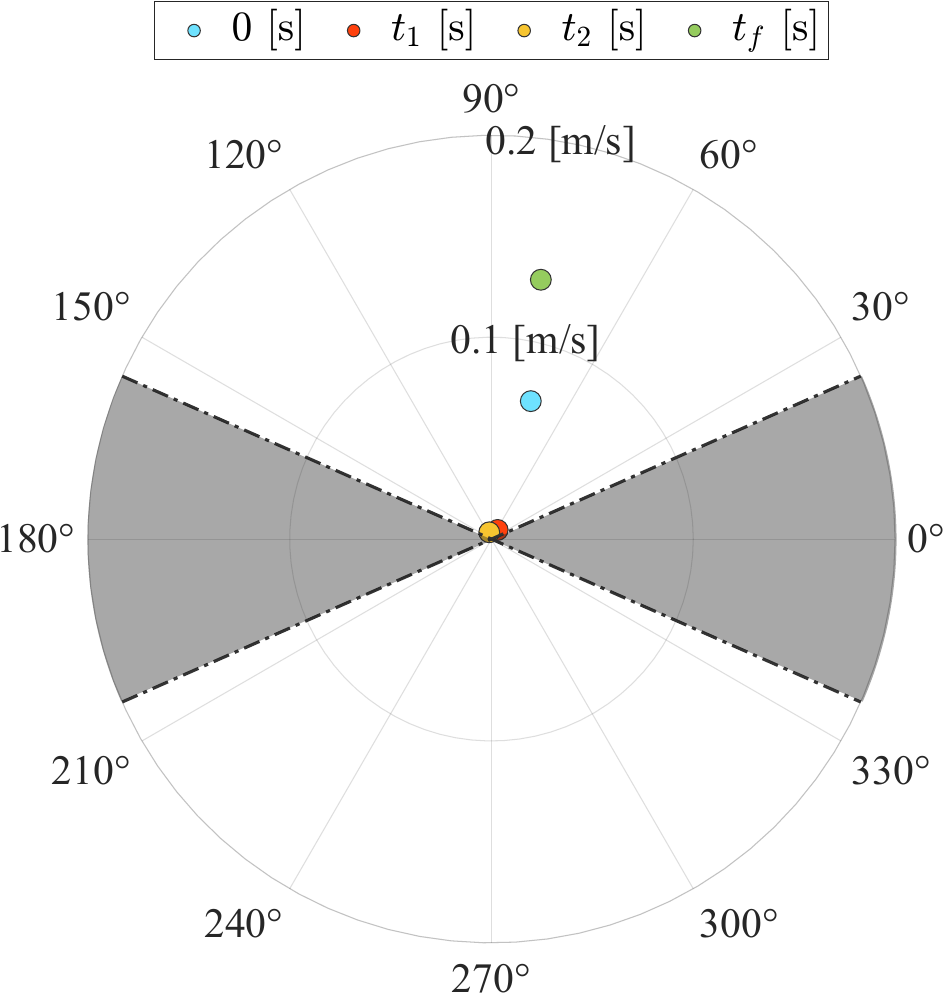}
         \vspace{1cm}  \caption{}
        \label{fig:controlconsts}
     \end{subfigure}
        \hfill
     \begin{subfigure}[b]{0.7\textwidth}
         \centering
         \includegraphics[width=\textwidth]{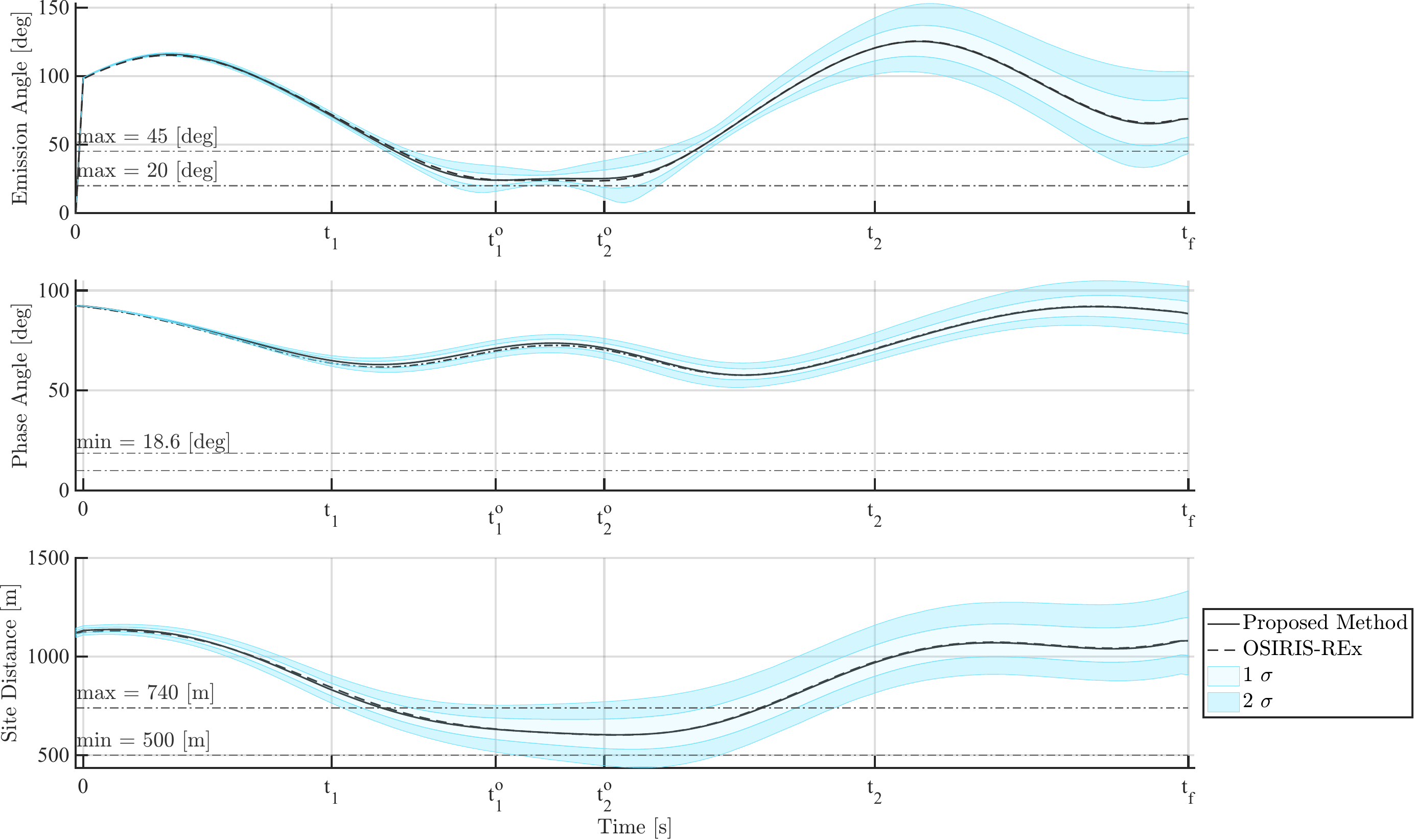}
    \caption{}
         \label{fig:result_angles}
     \end{subfigure}
        \caption{The converged solution given by the proposed planner. Figure (a) displays the norm of the input and the angle between the input and the vector from the spacecraft to the Sun. The gray highlighted area represents the keep-out zone indicated by Eq.~\eqref{eq:batterydeltavconsts}. Figure (b) shows the time histories of observation metrics, emission angle (top), phase angle (center), and site distance (bottom). Each figure represents the worst $1\sigma$ and $2\sigma$ dispersion of each (light blue and blue highlighted areas). Dotted horizontal lines represent the boundaries of each constraint.}
        \label{fig:resultsdawda}
\end{figure}

We begin by providing the performance of the proposed framework, whose 2D projected trajectories are shown in Figure~\ref{fig:result_1}. There are several notable observations from the results.
Firstly, we see in this figure that the solution state profile nominally satisfies Eq.~\eqref{eq:safetyconsts}.
Second, the primary difference between the trajectory given by the proposed planner and the as-flown~\ac{OREX} trajectory are seen in the center figure, which shows the trajectories projected on the x-y plane of the J2000 frame.
This difference results in the difference in emission angle histories, as depicted in Figure~\ref{fig:result_angles}.
Third, the chemical thruster fires mainly at the beginning and the end of the phase, which is similar to the classical bang-bang min-fuel solution.
However, two additional firings are executed to modify the trajectory to satisfy all the constraints under uncertainty. 
The control history is shown in Figure~\ref{fig:controlconsts}, which demonstrates that Eq.~\eqref{eq:batterydeltavconsts} and Eq.~\eqref{eq:controlconsts} are satisfied at every firing. 
Figure~\ref{fig:result_angles} shows the observation metrics histories, emission angle, phase angle, and site distance.
We observe in this figure that the solution nominal trajectory satisfies all of observation constraints during the observation time window.
We also note that the emission angle profile of the proposed planner is slightly lifted at the end of the observation time window, compared to the emission angle profile of the as-flown OSIRIS-REx trajectory.
This suggests the proposed planner modified the as-flown trajectory to improve observation possibility under uncertainties, which will be discussed later in detail. 
\begin{figure}[t!]
     \centering
      \begin{subfigure}[b]{0.3\textwidth}
         \centering
         \includegraphics[width=\textwidth]{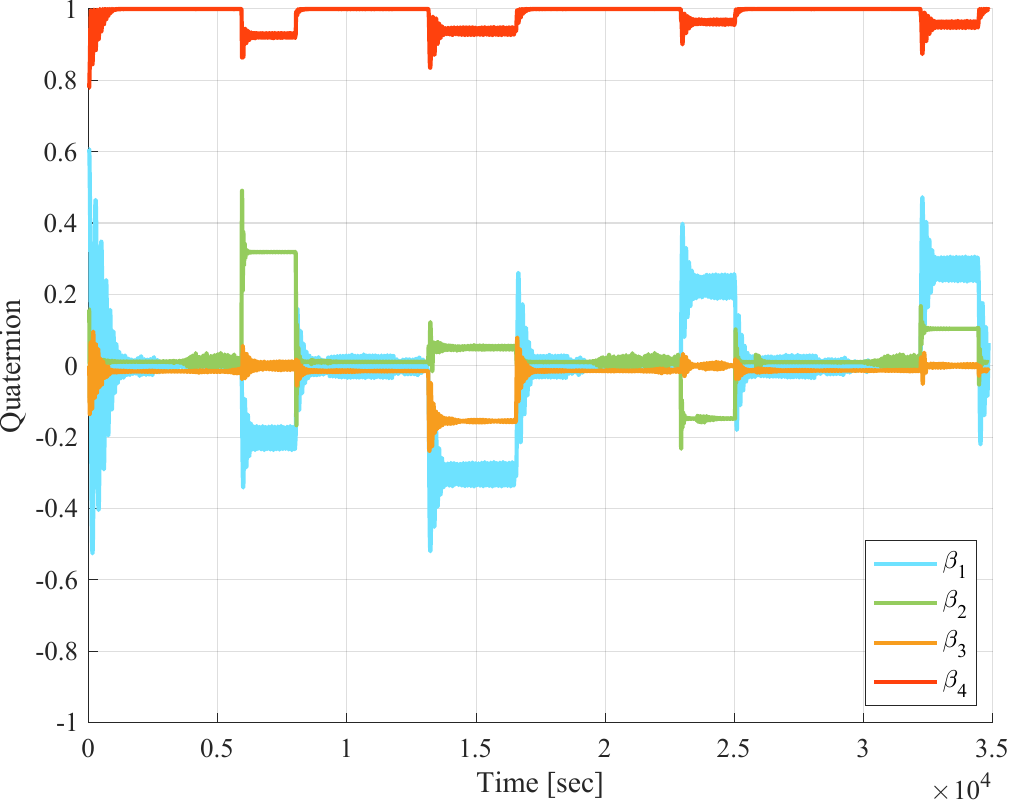}
            \caption{}
     \end{subfigure}
    \hfill
     \begin{subfigure}[b]{0.3\textwidth}
         \centering
         \includegraphics[width=\textwidth]{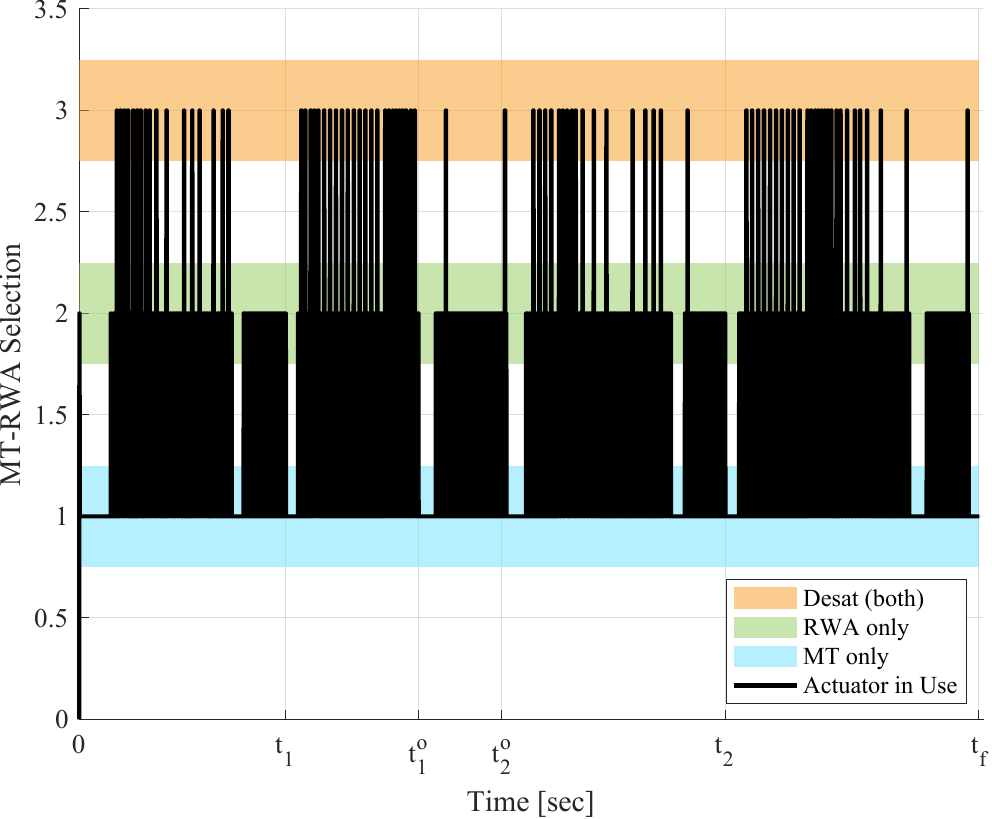}
         \caption{}
         \label{fig:muscat_attitude}
     \end{subfigure}
    \hfill
     \begin{subfigure}[b]{0.3\textwidth}
         \centering
         \includegraphics[width=\textwidth]{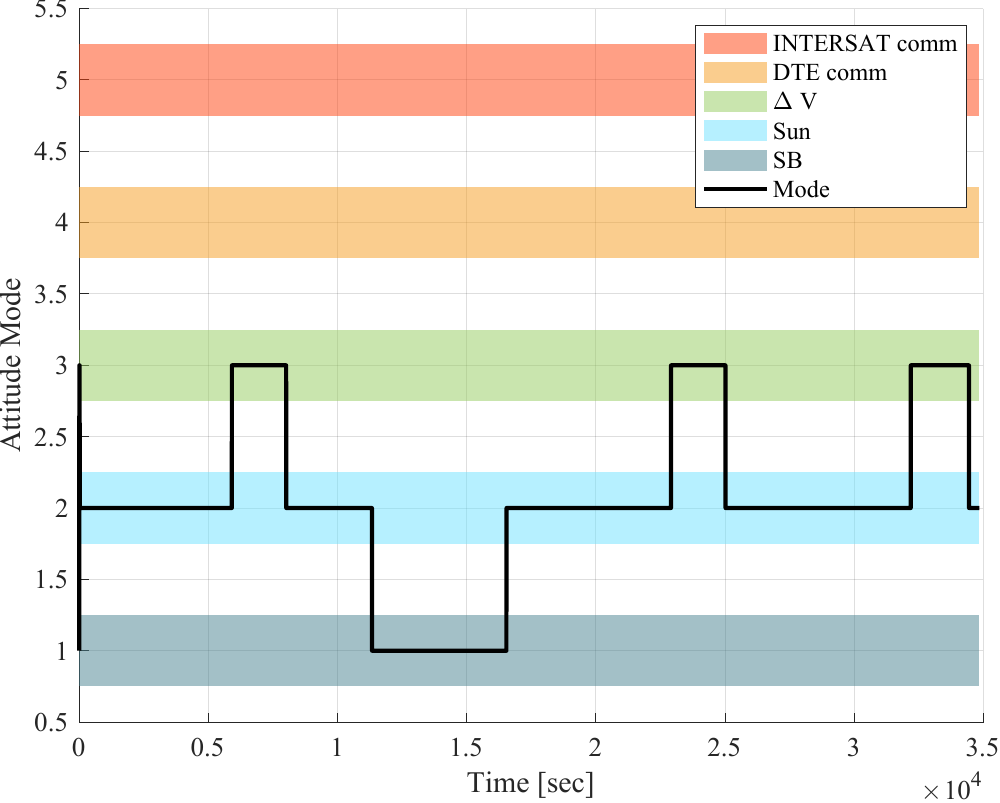}
         \caption{}
     \end{subfigure}
         \caption{Evolution of the spacecraft attitude for the converged trajectory. We note that the attitude of the spacecraft is controlled by an attitude GNC subsystem after pre-scheduling mode profiles. Figure (a) displays attitude time histories encoded by quaternions, where the subsystem steadily follow desired attitude time histories to realize pre-scheduled modes (shown in Figure (c)) with a combination of reaction wheels and multiple thrusters (shown in Figure (b)).
         }
        \label{fig:muscat_attitude_mode}
\end{figure}

\begin{figure}[t!]
     \centering
      \begin{subfigure}[b]{0.33\textwidth}
         \centering
         \includegraphics[width=\textwidth]{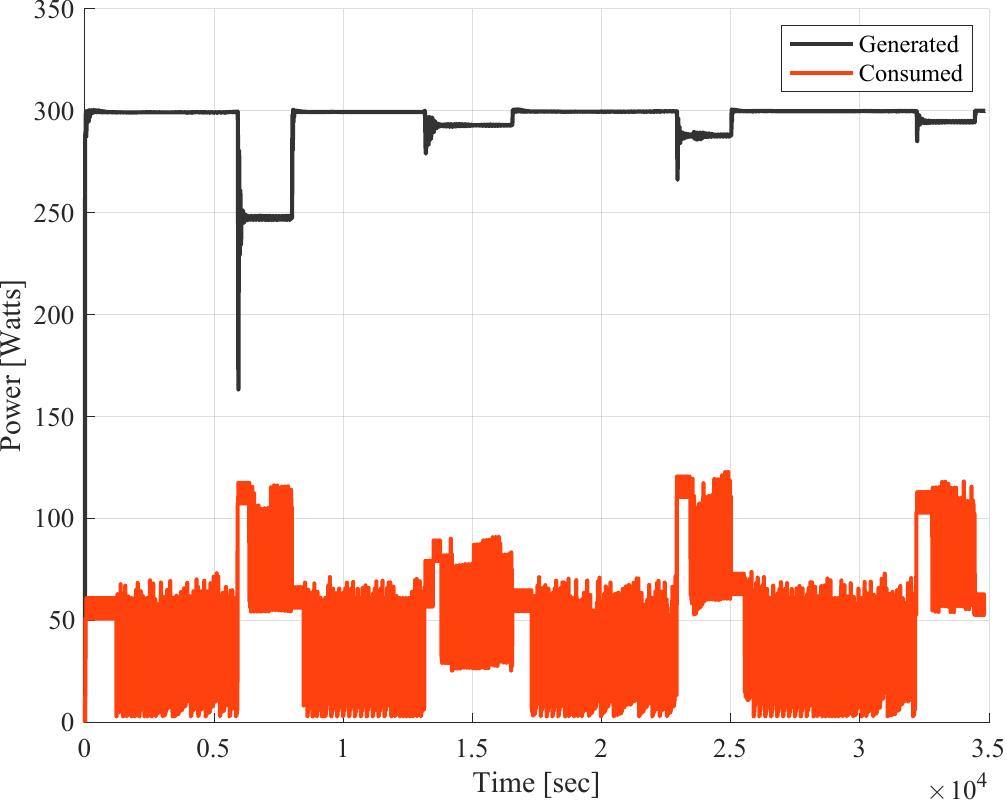}
         \caption{}
         \label{fig:muscat_battery_state}
     \end{subfigure}
    \hfill
     \begin{subfigure}[b]{0.33\textwidth}
         \centering
         \includegraphics[width=\textwidth]{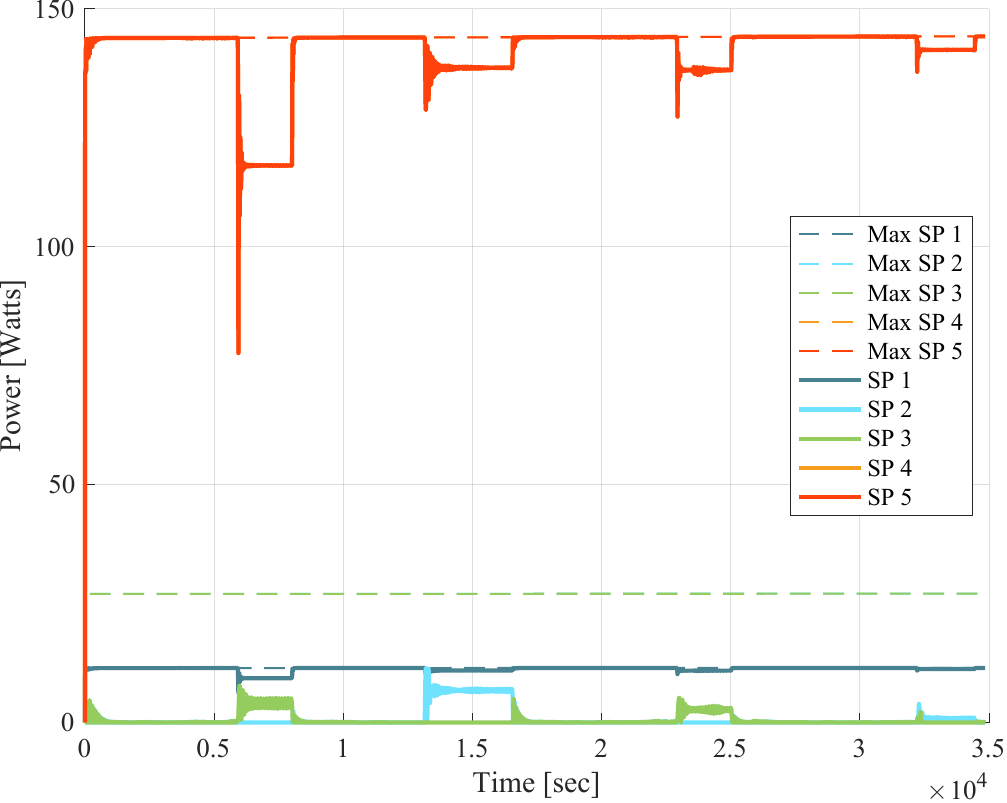}
         \caption{}
        \label{fig:muscat_battery_sps}
     \end{subfigure}
    \hfill
     \begin{subfigure}[b]{0.33\textwidth}
         \centering
         \includegraphics[width=\textwidth]{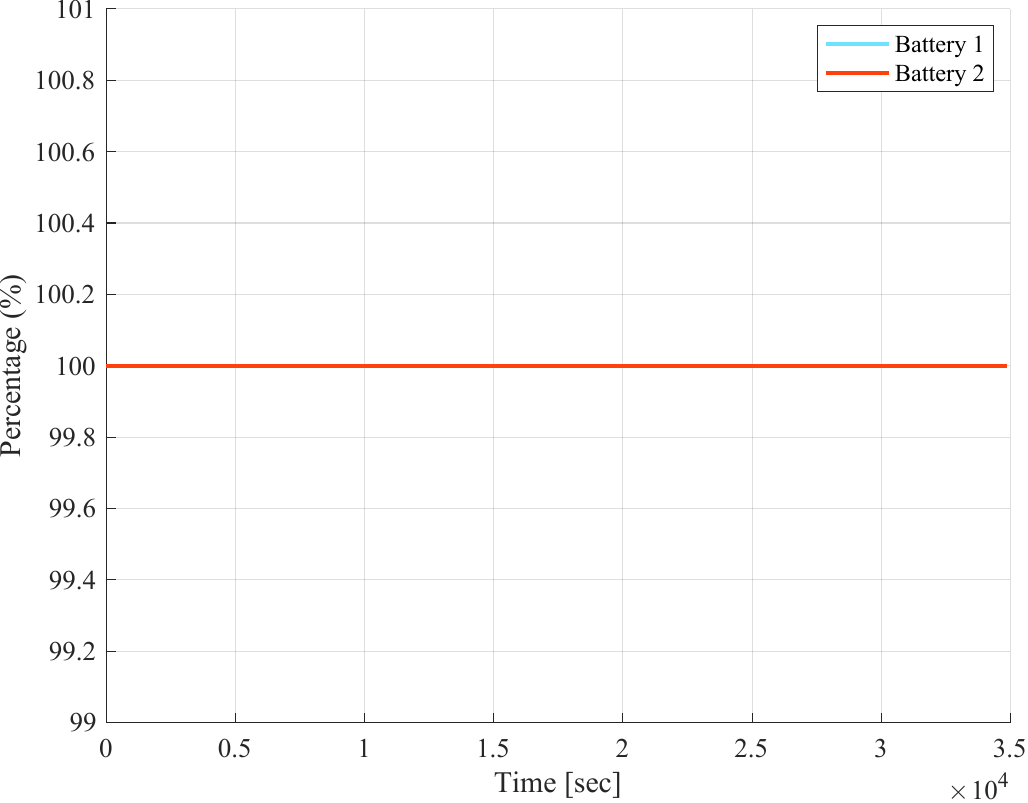}
         \caption{}
          \label{fig:muscat_battery_soc}
     \end{subfigure}
              \caption{ Battery generation and consumption time histories (a), power generation time histories from each of solar panels along with the maximum power generation for each (dotted line) (b), and the time histories of battery state of charges (c). Figure (a) displays that the power generation has been about three times larger than the power consumption throughout the reconnaissance phase. As a result, the states of charges have been kept to almost $100\%$ throughout the phase, as seen in Figure (c).}
        \label{fig:muscat_battery}
\end{figure}

\begin{table}[t!]
    \centering
    \begin{tabular}{ccccc}
    \hline
                     &                    & \multicolumn{3}{c}{\textbf{Observation Time}} \\\cline{3-5}
    \textbf{Methods} &  $\mathbf{t_2^o}$ - $\mathbf{t_1^o}$ & \textbf{Without Uncertainty} & \textbf{Mean} & \textbf{Worst} $\mathbf{1\sigma}$\\ \hline
    OSIRIS-REx       &  $\SI{3397}{s}$    & $\SI{3397}{s}$ & $\SI{2930}{s}$  & $\SI{2785}{s}$ \\
    Proposed Method with Warm Start  &  $\SI{3397}{s}$    & $\SI{3397}{s}$ & $\SI{3043}{s}$ & $\SI{3017}{s}$ \\ 
    Proposed Method with Cold Start &  $\SI{3397}{s}$    & $\SI{3397}{s}$ & $\SI{3271}{s}$ & $\SI{3397}{s}$ \\ \hline
    \end{tabular}
    \caption{Monte Carlo simulation results using three different methods, the current practice used in the OSIRIS-REx mission, the proposed trajectory planner with warm start (using the as-flown trajectory as an initial guess for the planner), and the proposed trajectory planner with cold start (using the linear interpolation between $x_0$ and $x_f$ as an initial guess for the planner). In either way of starting, the proposed method outperforms the current practice.}
    \label{tab:punchline}
\end{table}

\begin{figure}[t!]
    \centering    \includegraphics[width=0.6\textwidth]{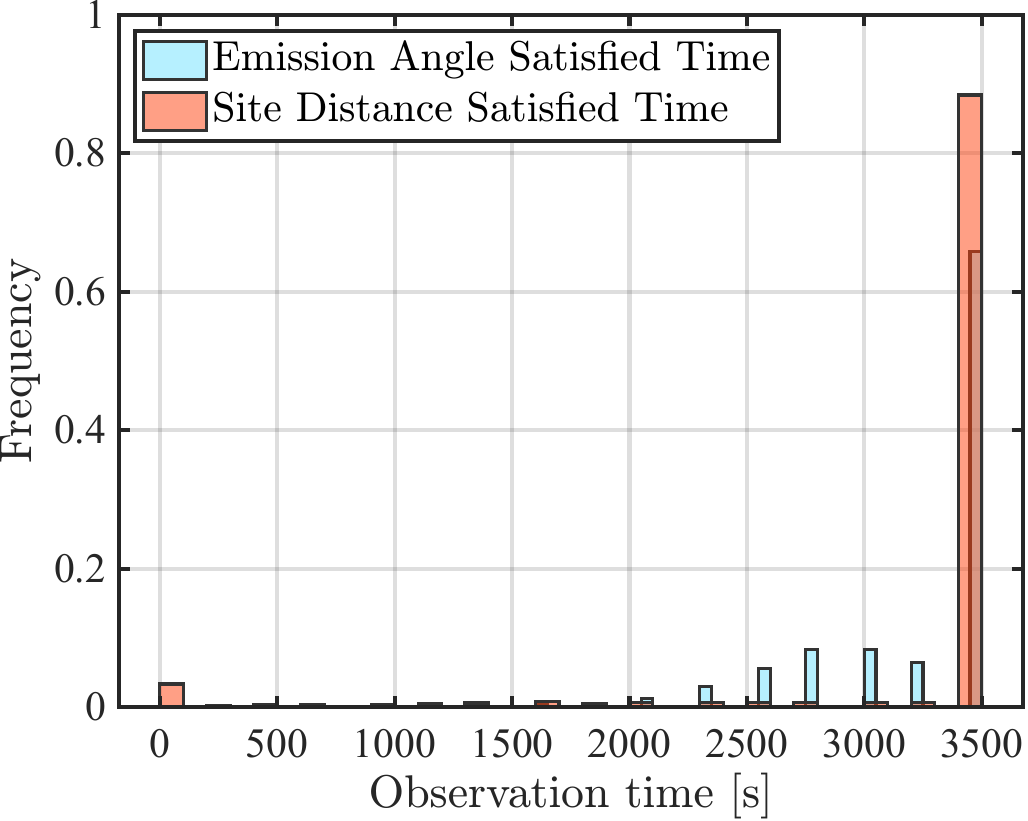}
    \caption{Resulting observation time distributions in which emission angle constraints (blue) or site distance constraints are satisfied (red).}
    \label{fig:histogram}
\end{figure}

In order to evaluate the proposed approach to decouple attitude and battery from the planner, we simulated the solution control profile in~\ac{MuSCAT}.
Figure~\ref{fig:muscat_attitude_mode} displays the mode selection profile and the resulting the attitude state profile, which is navigated and controlled by the attitude GNC subsystem simulated in~\ac{MuSCAT}.
We observe that the the current configuration of actuators (\ie{}, four reaction wheels and multiple thrusters) steadily track attitude modes indicated by the proposed planner, including delta-V pointing, small-body pointing, and recharging. 
Figure~\ref{fig:muscat_attitude} shows that the attitude GNC subsystem successfully use those two actuators depending on the desired torque, while handling with reaction wheels desaturation procedure. 
There are several notable observations from the results.
First, fixing the attitude to maximize the power generation increases the angular velocity of the reaction wheels (RW), which leads to several desaturations.
Next, changes in attitude associated with delta-V and observations require a significant amount of torque compared to maintaining attitude, so multiple thrusters are primarily used. 
Overall, we can conclude that our decoupling approach for the sake of numerical stability is reasonable for this mission concept. Moreover, the results of battery profiles simulated on~\ac{MuSCAT} are observed in Figure~\ref{fig:muscat_battery}. 
The battery consumption is not constant for each control mode. 
This is because~\ac{MuSCAT} simulates the uncertainties included in battery consumption / generation. Figure~\ref{fig:muscat_battery_state} plots the power consumption and generation profile. As discussed previously, battery consumption is increased when warming a chemical thruster up before firings. We see that the power generation consistently exceeds the power consumption with a margin throughout the phase. Figure~\ref{fig:muscat_battery_sps} shows the power generations from five different solar panels mounted on the spacecraft. This indicates that the attitude GNC subsystem is functioning correctly and the solar panel with the highest power generation, SP5, has been aimed towards the sun during the most of the reconnaissance phase. As a result, the battery~\ac{SoC} has been kept to $100\%$ throughout the reconnaissance phase, as seen in Figure~\ref{fig:muscat_battery_soc}.
Consequently, we can also conclude that the pre-scheduled mode profile is feasible in terms of the battery~\ac{SoC}.

Here, we provide a statistic analysis on the performances of the proposed approach and the as-flown trajectory in the~\ac{OREX} mission.
Our analysis consists of Monte Carlo simulations with $10,000$ samples for each approach, where randomly sampled control and initial state uncertainties are propagated around the corresponding nominal trajectories.
With respect to each sampled path, we measure time duration in which all observation constraints are satisfied.
Figure~\ref{fig:result_1} also shows Monte Carlo simulation results for the solution trajectory given by the proposed method.
We note that the distance from the small body center of mass at the worst $\SI{0.003}{}$ ($=3\sigma$) quantile was $\SI{667.24}{m}$; thus, the collision avoidance constraint was easily satisfied. Further, we see in Figure~\ref{fig:result_angles} that the site distance constraint and the emission angle were violated at the worst $2\sigma$, which leads to a limit on the observation time. Figure~\ref{fig:histogram} presents the distribution of durations in which each observation constraint is satisfied.
The emission angle is more prone to violating constraints compared to altitude.
Such violations, although frequent, result in a relatively minor reduction in observation time, typically not exceeding $\SI{1000}{s}$ per occurrence. On the other hand, although violations related to altitude are less frequent, they present a more severe problem when they occur.
A single incident of altitude non-compliance leads to zero effective observation time.
The other observation constraint, the phase angle constraint, is satisfied even at the worst $2\sigma$.
Given that all observation constraints are satisfied at the worst $1 \sigma$, we conclude that the converged solution is feasible regarding stochastic constraints.

In order to evaluate the performances of the proposed planner and the current practice used in the~\ac{OREX}, Table~\ref{tab:punchline} provides statistical analysis on observation time.
Both the proposed trajectory and the as-flown Reconnaissance B trajectory for~\ac{OREX} trajectory were designed to satisfy all of the observation constraints during $t_1^o \leq t \leq t_2^o$. Therefore, the resulting observation times without uncertainties are equal to the maximum observation duration. 
However, the observation time given by the proposed planner is less sensitive than that of the as-flown trajectory. 
Additionally, the proposed planner with cold start achieves a better result than the proposed planner with warm start, indicating that the achieved result is not due to the quality of the initial guess.
Consequently, we can conclude that the proposed planner increases the probability of successfully acquiring observations compared to the as-flown~\ac{OREX} trajectory. 

Overall, we conclude that these numerical results show that our proposed framework has two advantages: making the trajectory less sensitive to uncertainty, and efficient computation of solutions offered by the off-the-shelf nonlinear optimization solvers.

\section{Conclusion}
We presented a framework for autonomous trajectory planning for the reconnaissance phase of a small-body exploration mission subject to initial state and environmental uncertainties.
Our approach involves modeling the trajectory planning problem as a stochastic trajectory optimization problem and efficiently solving it using off-the-shelf nonlinear optimization solvers.
We focused on the proposed~\ac{DARE} mission as our case study and, importantly, formulated our modeling assumptions by rigorously validating them in the~\ac{MuSCAT} simulation framework.
Finally, Monte Carlo simulations of our proposed approach demonstrate that the autonomous trajectory planner outperforms the state-of-the-practice approaches for mission trajectory planning. 
In the future, we plan to improve the computation times of our planner by further optimizing our code for on-board use and extend the scope of our optimal control problem to also consider firing timings and the number of firings.
\textcolor{black}{
A shortcoming of the proposed framework is its potential failure to converge to a feasible solution, particularly with respect to covariance matrices.
While Monte Carlo simulations still demonstrate the efficacy of the framework in handling uncertainty and risk, a future research entails updating the framework to ensure the feasibility of the deterministic nominal trajectory.}
Through this effort, our results demonstrate the ability for such autonomous trajectory planners to consider the rich set of mission system and safety constraints that renders the trajectory design problem challenging.
For proposed missions such as~\ac{DARE} and other spacecraft missions, the inclusion of such autonomous trajectory planners can greatly simplify mission operations and pave the way for greater scientific returns and exploration.
To this end, we believe that the development and validation of the trajectory planning framework presented here serves as a key enabling bridge towards fully autonomous spacecraft missions.


\section*{Acknowledgments}
The research was carried out at the Jet Propulsion Laboratory, California Institute of Technology, under a contract with the National Aeronautics and Space Administration (80NM0018D0004). © 2024. All rights reserved. The lead author Kazuya Echigo's time was funded in-part by Air Force Office of Scientific Research grant FA9550-20-1-0053.

\bibliography{project}

\end{document}